\documentclass[nofootinbib,prd,aps]{revtex4}

\usepackage{graphicx}
\usepackage{dcolumn}
\usepackage{bm}

\def \be{\begin{equation}}
\def \ee{\end{equation}}

\begin{document}
\voffset = 0.3 true in

\title{Dynamic Properties of Charmonium}

\author{Olga Lakhina}
\affiliation{
Department of Physics and Astronomy, University of Pittsburgh,
Pittsburgh PA 15260}

\author{Eric S. Swanson}
\affiliation{
Department of Physics and Astronomy, University of Pittsburgh,
Pittsburgh PA 15260}

\begin{abstract}
\vskip .3 truecm
Nonrelativistic quark models of charmonia are tested by comparison of theoretical
charmonium decay constants, form factors, and $\gamma\gamma$ widths with experiment 
and lattice gauge computations. The importance of relativistic effects, a running coupling,
and the correct implementation of bound state effects are demonstrated. We describe how 
an improved model and computational techniques resolve several 
outstanding issues in previous nonrelativistic quark models such as 
the use of 
`correction' factors in quark model form factors, artificial energy prescriptions in decay
constant calculations, and {\it ad hoc} phase space modifications. We comment on 
the small experimental value of $f_{\psi''}$ and
the D-wave component of the $J/\psi$. Decay constants and $\gamma\gamma$ widths for bottomonium
are also presented.
\end{abstract}

\maketitle

\section{Introduction}  

New spectroscopy from the B factories and the advent of CLEO-c and the BES upgrade have
led to a resurgence of interest in charmonia. Among the new developments are the discovery
of the $\eta_c'$ and $h_c$ mesons and the observation of the enigmatic $X(3872)$ and
$Y(4260)$ states at Belle\cite{HHreview}. Furthermore, lattice gauge theory is now able
to produce reasonably accurate measurements of charmonia masses and form factors\cite{JLlatt}.
It is thus opportune to re-examine constituent quark model predictions of charmonia properties
in an attempt to refine current models, test quark models in new regimes, and look for the
expected failure of these models. 

It is evident that the quality of spectra is only a rough indication of model efficacy. Thorough
tests of models requires probing quark dynamics in different regimes. We shall pursue this by
computing charmonia observables
such as decay constants, elastic and transition form factors, and $\gamma\gamma$ decay rates.
This investigation is therefore
complementary to that of Ref. \cite{BGS}, which examined spectra, electromagnetic transitions,
and strong decay rates. We remark that the latter is a nonperturbative process which 
requires further modelling
in contrast to the observables computed here, which are driven by well-defined electroweak currents.

In the following we will demonstrate that agreement with experimental charmonium decay constants
requires a weakening of the short range quark interaction with respect to the standard 
Coulomb interaction.
This weakening is in accord with the running coupling of perturbative QCD and eliminates the
need for an artificial energy dependence that was introduced by Godfrey and Isgur\cite{GI} to
fit experimental decay constants.

Single quark elastic and transition form factors are considered in Sections \ref{sqffSect} 
and \ref{tffSect}. The agreement with recent lattice computations is very good, but requires
that the standard nonrelativistic reduction of the current not be made and that the running coupling
described above be employed. As will be shown, this obviates the need for the phenomenological
$\kappa$ factor 
introduced for electroweak decays in the ISGW model\cite{ISGW}. Analogous results for bottomonium
are presented in Appendix \ref{bottomApp}.

Section \ref{ggSect} analyzes the two photon decays of charmonia. We argue that this decay 
should be described in terms of bound state perturbation theory and that it is therefore a 
convolution of form factors and decay constants. In contrast with traditional approaches, 
the resulting computations are in good
agreement with experiment  and improve the agreement with low energy theorems.
This permits the elimination of an artificial mass dependence employed in Ref. \cite{GI} in an
attempt to improve agreement with experiment.

\section{Nonrelativistic Charmonium Structure}  

We adopt the standard practice of describing charmonia with nonrelativistic kinematics, a
central confining potential, and order $v^2/c^2$ spin-dependent interactions. Thus $H =  2m + P_{rel}^2/2\mu + V_C
+ V_{SD}$ where 

\be
V_C(r) = 
-\frac{4}{3}\frac{\alpha_C}{r} + br,
\label{vcEq}
\ee
and
\be
V_{SD}(r) = 
\frac{32\alpha_H\pi e^{-k^2/4\sigma^2}}{9 m_q^2 }
\vec{S}_q \cdot \vec{S}_{\bar q}  + \Big(\frac{2\alpha_s}{r^3}-\frac{b}{2r}\Big)\frac{1}{m_q^2}\vec L \cdot \vec S + \frac{4\alpha_s}{m_q^2 r^3} T,
\label{vsdEq}
\ee
where $3T = 3\hat r \cdot S_q \hat r \cdot \vec S_{\bar q} - \vec S_q \cdot \vec S_{\bar q}$.
The strengths of the Coulomb and hyperfine interactions have been taken as separate parameters.
Perturbative gluon exchange implies that $\alpha_C = \alpha_H$ and we find that the fits
prefer the near equality of these parameters. 

As will be described below, the observables considered here require a weaker ultraviolet 
interaction than
that of Eq. \ref{vcEq}. We therefore introduce a running coupling that recovers the perturbative
coupling of QCD but saturates at a phenomenological value at low momenta:

\be
\alpha_C \to \alpha_C(k)=\frac{4\pi}{\beta_0\log\left({e^{\frac{4\pi}{\beta_0\alpha_0}}+\frac{k^2}{\Lambda^2}}\right)}
\label{alRunEq}
\ee
with $\beta_0$ taken to be 9.
One can identify the parameter $\Lambda$ with $\Lambda_{QCD}$ because $\alpha_C(k)$ approaches
the one loop running constant of QCD. However, this parameter will also be fit to experimental
data in the following (nevertheless, the resulting preferred value is reassuringly close to 
expectations). Parameters and details of the fit are presented in Appendix \ref{modelsApp}.


The resulting low lying spectra are presented in Table \ref{spectrumTab}. The first
column presents the results of the `BGS'  model\cite{BGS}, which was tuned to the available
charmonium spectrum. The second and third columns, labelled BGS+log, makes the replacement of Eq. \ref{alRunEq};
the parameters have not been retuned. One sees that the $J/\psi$ and $\eta_c$ masses have been
raised somewhat and that the splitting has been reduced to 80 MeV. Heavier states have only
been slightly shifted. It is possible to fit the $J/\psi$ and $\eta_c$ masses by adjusting
parameters, however this tends to ruin the agreement of the model with the excited states. We
therefore choose to compare the BGS and BGS+log models without any further adjustment to the
parameters.
A comparison with other models and lattice gauge theory can be found in
Ref. \cite{HHreview}.

\begin{table}[!h]
\caption{\label{mass} Spectrum of $c\bar{c}$ mesons (GeV).}
\begin{center}
\begin{tabular}{c|cccc}
\hline
state & BGS & BGS log  & BGS log & experiment \\
      &     & $\Lambda = 0.25$ GeV & $\Lambda = 0.4$ GeV & \\
\hline
\hline
$\eta_c(1^1S_0)$    &2.981 &3.088  &3.052  &2.979\\
$\eta_c(2^1S_0)$    &3.625 &3.669  &3.655  &3.638\\
$\eta_c(3^1S_0)$    &4.032 &4.067  &4.057  &-    \\
$\eta_c(4^1S_0)$    &4.364 &4.398  &4.391  &-    \\
$\eta_{c2}(1^1D_2)$ &3.799 &3.803  &3.800  &-    \\
$\eta_{c2}(2^1D_2)$ &4.155 &4.158  &4.156  &-    \\
$J/\psi(1^3S_1)$    &3.089 &3.168  &3.139  &3.097\\
$\psi(2^3S_1)$      &3.666 &3.707  &3.694  &3.686\\
$\psi(3^3S_1)$      &4.060 &4.094  &4.085  &4.040\\
$\psi(4^3S_1)$      &4.386 &4.420  &4.412  &4.415\\
$\psi(1^3D_1)$      &3.785 &3.789  &3.786  &3.770\\
$\psi(2^3D_1)$      &4.139 &4.143  &4.141  &4.159\\
$\psi_2(1^3D_2)$    &3.800 &3.804  &3.801  &-    \\
$\psi_2(2^3D_2)$    &4.156 &4.159  &4.157  &-    \\
$\psi_3(1^3D_3)$    &3.806 &3.809  &3.807  &-    \\
$\psi_3(2^3D_3)$    &4.164 &4.167  &4.165  &-    \\
$\chi_{c0}(1^3P_0)$    &3.425 &3.448  &3.435  &3.415\\
$\chi_{c0}(2^3P_0)$    &3.851 &3.870  &3.861  &-    \\
$\chi_{c0}(3^3P_0)$    &4.197 &4.214  &4.207  &-    \\
$\chi_{c1}(1^3P_1)$    &3.505 &3.520  &3.511  &3.511\\
$\chi_{c1}(2^3P_1)$    &3.923 &3.934  &3.928  &-    \\
$\chi_{c1}(3^3P_1)$    &4.265 &4.275  &4.270  &-    \\
$\chi_{c2}(1^3P_2)$    &3.556 &3.564  &3.558  &3.556\\
$\chi_{c2}(2^3P_2)$    &3.970 &3.976  &3.972  &-    \\
$\chi_{c2}(3^3P_2)$    &4.311 &4.316  &4.313  &-    \\
$h_c(1^1P_1)$       &3.524 &3.536  &3.529  &-    \\
$h_c(2^1P_1)$       &3.941 &3.950  &3.945  &-    \\
$h_c(3^1P_1)$       &4.283 &4.291  &4.287  &-    \\
\hline
\hline
\end{tabular}
\end{center}
\label{spectrumTab}
\end{table}

%

As has been stressed above, the spectrum is not a particularly robust test of model reliability 
because it only probes gross
features of the wavefunction. Alternatively, observables such as strong and electroweak decays probe
different wavefunction  momentum scales. For example, decay constants are short distance observables
while strong and radiative transitions test intermediate scales.  Thus the latter do not add
much new information unless the transition occurs far from the zero recoil point. In this case
the properties of boosted wavefunctions and higher momentum components  become important. 
We choose to compute charmonium decay constants, elastic and transition form factors, and $\gamma\gamma$
decays in the following.

\section{Charmonium Decay Constants}  
\label{fSect}

Leptonic decay constants are a simple probe of the short distance structure of hadrons and therefore
are a useful observable for testing quark dynamics in this regime. Decay constants are computed
by equating their field theoretic definition with the analogous quark model definition. This 
identification is rigorously valid in the nonrelativistic and weak binding limits where
quark model state vectors form good representations
of the Lorentz group\cite{ISGW,HI}. The task at hand is to determine the reliability of the computation 
away from these limits.

The method is illustrated with the vector charmonium decay constant $f_V$,  which is defined by

\be
m_V \, f_V \, \epsilon^{\mu}= \langle 0|\bar{\Psi}\gamma^{\mu}\Psi|V\rangle
\label{VdecayEq}
\end{equation}
where $m_V$ is the vector meson mass and $\epsilon^{\mu}$ is its polarization vector. 
The decay constant is computed in the conceptual weak binding and nonrelativistic limit
of the quark model and is assumed to be accurate away from these limits. One thus employs the
quark model state:

\begin{equation}
|V(P)\rangle = \sqrt{\frac{2 E_P}{N_c}} 
\chi^{SM_S}_{s\bar{s}}\, 
\int\frac{d^3k\, d^3\bar{k}}{(2\pi)^3}\Phi\left(\frac{m_{\bar{q}}\vec{k}-m_q\vec{\bar{k}}}{m_{\bar{q}}+m_q}\right)
\delta^{(3)}(\vec{k}+\vec{\bar{k}}-\vec{P}) b^\dag_{ks} d^\dag_{\bar{k}\bar s}|0\rangle.
\label{VmesonEq}
\end{equation}
The decay constant is obtained by
computing the spatial matrix element of the current in the
vector center of mass frame (the temporal component is trivial) and yields

\begin{equation}
f_V=\sqrt{\frac{N_c}{m_V}}\int\frac{d^3k}{(2\pi)^3} \Phi(\vec{k})
\sqrt{1+\frac{m_q}{E_k}} \sqrt{1+\frac{m_{\bar{q}}}{E_{\bar{k}}}}
\left(1+\frac{k^2}{3(E_k+m_q)(E_{\bar{k}}+m_{\bar{q}})}\right).
\label{relfVEq}
\end{equation}
The nonrelativistic limit is proportional to the meson wave function at the origin
\begin{equation}
f_V
= 2\sqrt{\frac{N_c}{m_V}} \tilde{\Phi}(r=0);
\label{nonrelfVEq}
\end{equation}
which recovers the well-known result of van Royen and Weisskopf\cite{vRW}.

Similar results hold for other charmonia that couple to electroweak currents.
A summary of the results for a variety of models
are presented in Table \ref{dcTab}.
The expressions used to compute the table entries and the data used to extract the experimental
decay constants are collected in
Appendix \ref{DecayConstantsApp}.

\begin{table}[!h]
\caption{Charmonium Decay Constants (MeV).}
\label{dcTab}
\begin{tabular}{c|cccccc}
\hline
Meson & BGS NonRel & BGS Rel & BGS log & BGS log & lattice & experiment \\
      &            &         &  $\Lambda= 0.4$ GeV & $\Lambda = 0.25$ GeV & & \\   
\hline
\hline
$\eta_c$                       &795    &493   &424   &402      &$429\pm 4\pm 25$  &$335\pm 75$\\
$\eta'_c$                      &477    &260   &243   &240      &$56\pm 21\pm 3$   &\\
$\eta''_c$                     &400    &205   &194   &193      &                  &\\
$J/\psi$                       &615    &545   &423   &393      &$399\pm 4$        &$411\pm 7$\\
$\psi'$                        &431    &371   &306   &293      &$143\pm 81$       &$279 \pm 8$\\
$\psi''$                       &375    &318   &267   &258      &                  &$174\pm 18$\\
$\chi_{c1}$                       &239    &165   &155    &149       &                  &\\
$\chi'_{c1}$                      &262    &167   &157   &152      &                  &\\
$\chi''_{c1}$                     &273    &164   &155   &151      &                  &\\
\hline
\end{tabular}
\end{table}

The second column shows results of the nonrelativistic computation  (Eq. \ref{nonrelfVEq}) with 
wavefunctions determined
in the Coulomb+linear BGS model.  A clear trend is evident as all predictions are approximately
a factor of two larger than experiment (column seven). Using the full spinor structure (column three)
improves agreement with experiment substantially, but still yields predictions which are roughly 30\%
too large.  At this stage the lack of agreement must be ascribed to strong dynamics, and this 
motivated the running coupling model specified above. The fourth and fifth columns give the 
results obtained from this model.
It is apparent that the
softening of the short range Coulomb potential induced by the running coupling brings the
predictions into very good agreement with experiment.

Column six lists the quenched lattice gauge computations of Ref. \cite{JLlatt}. The agreement with
experiment is noteworthy; however, the predictions for the $\eta_c'$ and
$\psi'$ 
decay constants  are much smaller than those of the quark model (and experiment in the case
of the $\psi'$). It is possible that this is due to excited state contamination in the 
computation of the mesonic correlators.

The good agreement between model and experiment has been obtained with a straightforward 
application of the
quark model. This stands in contrast to the methods adopted in Ref. \cite{GI} where the
authors insert arbitrary factors of $m/E(k)$ in the integrand in order to obtain agreement
with experiment (the extra factors of $m/E$ serve to weaken the integrand, approximating the
effect of the running coupling used here).

It is very difficult to obtain a value for $f_{\psi''}$ that is as small as experiment. 
Assuming that the experimental value is reliable it is possible that this difficulty points
to serious problems in the quark model. A simple mechanism for diminishing the decay constant
is via S-D wave mixing, because the D-wave decouples from the 
vector current. This mixing can be generated by the tensor interaction of Eq. \ref{vsdEq};
however, computations yield amplitude reductions of order 2\% -- too small to explain the
effect. Note that S-D mixing can also be created by transitions to virtual meson-meson pairs. 
Unfortunately,
evaluating this requires a reliable model of strong  Fock sector mixing and we do not pursue
this here.

A similar discussion holds for the $e^+e^-$ width of the $\psi(3770)$. Namely, the large decay constant
$f_{\psi(3770)} = 99 \pm 20$ MeV can perhaps be explained by mixing with nearby S-wave states. 
Again, the computed effect due to the tensor interaction is an order of magnitude too small and 
one is forced to look elsewhere (such as loop effects) for an explanation.

Attempts to compute Lorentz scalars such as decay constants or form factors in a noncovariant
framework are necessarily ambiguous.  As stated above, the results of a computation in the
nonrelativistic quark model are only guaranteed to be consistent in the weak binding limit.
However the accuracy of the quark model can be estimated by examining the decay constant dependence
on model assumptions.
For example,
an elementary aspect of covariance is that a single decay constant describes the vector (for example)
decay amplitude in all frames and for all four-momenta. Thus the decay constant computed from the
temporal and spatial components of the matrix element $\langle 0 | J_\mu|V\rangle$ should be equal.
As pointed out above, setting $\mu=0$ yields the trivial result $0=0$ in the vector rest frame.
However, away from the rest frame one obtains the result

\begin{equation}
f_V = \sqrt{N_c E(P)} \int {d^3 k \over (2\pi)^3} \Phi(k;P) {1 \over \sqrt{E(k+P/2)}\sqrt{E(k-P/2)}}
\frac{1}{2}\left( \frac{\sqrt{E(k+P/2)+m}}{\sqrt{E(k-P/2)+m}} + 
                  \frac{\sqrt{E(k-P/2)+m}}{\sqrt{E(k+P/2)+m}}  \right)
\end{equation}
or, in the nonrelativistic limit
\begin{equation}
f_V = {\sqrt{N_c M_V}\over m} \tilde\Phi(0).
\end{equation}
One sees that covariance is recovered in the weak binding limit where the 
constituent quark model is formally valid. 

\begin{figure}[h]
\includegraphics[angle=-90,width=9cm]{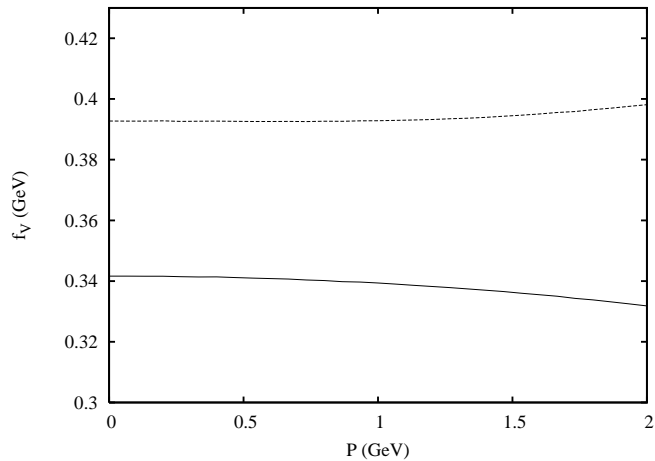}
\caption{Temporal (top line) and Spatial (bottom line) Vector Decay Constants in Various Frames.}
\label{covDecayConstantFig}
\end{figure}

Computations of the vector decay constant away from the weak binding limit and the rest frame are
displayed in Fig. \ref{covDecayConstantFig}. One sees a reassuringly weak dependence on the
vector momentum $P$. There is, however, a 13\% difference in the numerical value of the 
temporal and spatial decay constants, which may be taken as a measure of the reliability of the 
method.

\section{Single Quark Elastic Form Factors} 
\label{sqffSect}

Form factors are a powerful determinant of internal hadronic structure because the external current
momentum serves as a probe scale. And of course, different currents are sensitive to 
different properties
of the hadron. The simplest form factors are elastic (such as the pion electromagnetic form factor) and
it is therefore useful to examine these when tuning and testing models. 
Unfortunately elastic electromagnetic form factors are not observables for charmonia; however 
this is an area where lattice gauge theory can aid greatly in the development of models 
and intuition.  In particular, a 
theorist can choose to couple the external current to a single quark, thereby yielding
a nontrivial `pseudo-observable'. This has been done in Ref. \cite{JLlatt} and we 
follow their lead here 
by considering
the single-quark elastic electromagnetic form factors for pseudoscalar, scalar, 
vector, and axial vector charmonia. 

The technique used to compute the form factors is illustrated by considering the inelastic pseudoscalar
electromagnetic matrix element $\langle P_2 | J^\mu | P_1\rangle$, where $P$ refers to a pseudoscalar
meson.
The most general Lorentz covariant decomposition of this matrix element is 

\begin{equation}
\langle P_2(p_2)|\bar{\Psi}\gamma^{\mu}\Psi|P_1(p_1)\rangle = f(Q^2)\left((p_2+p_1)^{\mu}-\frac{M^2_2-M^2_1}{q^2}(p_2-p_1)^{\mu}\right)
\end{equation}
where conservation of the vector current has been used to eliminate a possible second invariant.
The argument of the form factor is chosen to be $Q^2 = - (p_2-p_1)_\mu (p_2-p_1)^\mu$.

Using the temporal component of the vector current and computing in the rest frame of the
initial meson yields

\begin{eqnarray}
f_{sq}(Q^2)&=&\frac{\sqrt{M_1E_2}}{(E_2+M_1)-\frac{M^2_2-M^2_1}{q^2}(E_2-M_1)}\nonumber\\
&\times&
\int \frac{d^3k}{(2\pi)^3}
\Phi(\vec{k})
\Phi^*\left(\vec{k}+\vec{q}\frac{\bar{m}_2}{m_2+\bar{m}_2}\right)
\sqrt{1+\frac{m_1}{E_k}}\sqrt{1+\frac{m_2}{E_{k+q}}}
\left(1+\frac{(\vec{k}+\vec{q})\cdot\vec{k}}
{(E_k+m_1)(E_{k+q}+m_2)}\right)
\label{ffFullEq}
\end{eqnarray}
The pseudoscalars are assumed to have valence quark masses $m_1, \bar m_1$ and $m_2, \bar m_2$ for $P_1$ and
$P_2$ respectively. The masses of the mesons are labelled $M_1$ and $M_2$.
The single quark elastic form factor can be obtained by setting $m_1= \bar m_1 = m_2 = \bar m_2$
and $M_1 = M_2$. In the nonrelativistic limit Eq. \ref{ffFullEq} reduces to the simple expression:

%

\begin{equation}
f_{sq}(Q^2)=
\int \frac{d^3k}{(2\pi)^3} \Phi(\vec{k}) \Phi^*\left(\vec{k}+\frac{\vec{q}}{2}\right).
\label{ff0Eq}
\end{equation}
In this case it is easy to see the normalisation condition $f_{sq}(\vec q =0) = 1$. This is also
true for the relativistic elastic single quark form factor of Eq. \ref{ffFullEq}. 

\begin{figure}[h]
\includegraphics[angle=-90,width=9cm]{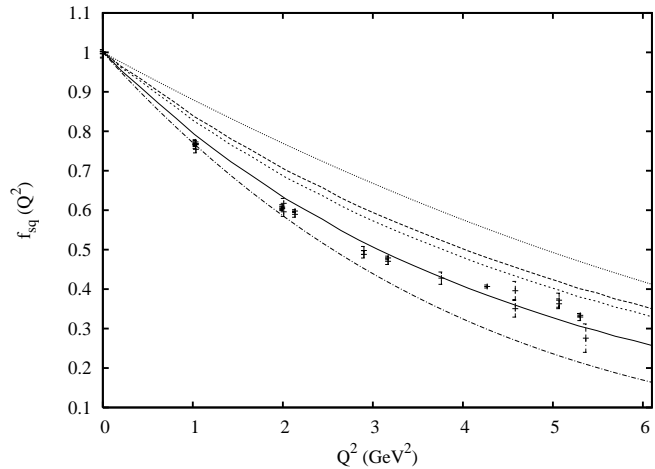}
\caption{The Single Quark $\eta_c$ Form-factor $f_{sq}(Q^2)$. From top to bottom the curves are
SHO, nonrelativistic BGS, relativistic BGS, BGS log, and ISGW.}
\label{etaFig}
\end{figure}

A variety of quark model computations of the $\eta_c$ single quark elastic form factor 
are compared to lattice results in Fig. \ref{etaFig}. 
It is common to use SHO wavefunctions when computing complicated matrix elements. 
The dotted curve displays the nonrelativistic form factor (Eq. \ref{ff0Eq}) with SHO
wavefunctions (the SHO scale is taken from Ref. \cite{cs}). Clearly the result is too hard with respect to the lattice.
This problem was noted by ISGW and is the reason they introduce a suppression
factor $\vec q \to \vec q/\kappa$. ISGW set $\kappa = 0.7$ to obtain agreement with the pion
electromagnetic form factor. The same procedure yields the dot-dashed curve in Fig. \ref{etaFig}. 
The results agrees well with lattice for small $Q^2$; thus, somewhat surprisingly, the {\it ad hoc}
ISGW procedure appears to be successful for heavy quarks as well as light quarks.

The upper dashed curve indicates that replacing SHO wavefunctions with full Coulomb+linear 
wavefunctions
gives a somewhat softer nonrelativistic form factor. The same computation with the relativistic 
expression (Eq. \ref{ffFullEq}), the lower dashed curve, yields a slight additional improvement. Finally,
the relativistic BGS+log single quark elastic $\eta_c$ form factor is shown as the 
solid line and is in remarkably good agreement with the lattice (it is worth stressing that
form factor data have not been fit). It thus appears that the ISGW
procedure is an {\it ad hoc} procedure to account for relativistic dynamics and deviations of simple
SHO wavefunctions from Coulomb+linear+log wavefunctions.

\begin{figure}[h]
\includegraphics[angle=-90,width=9cm]{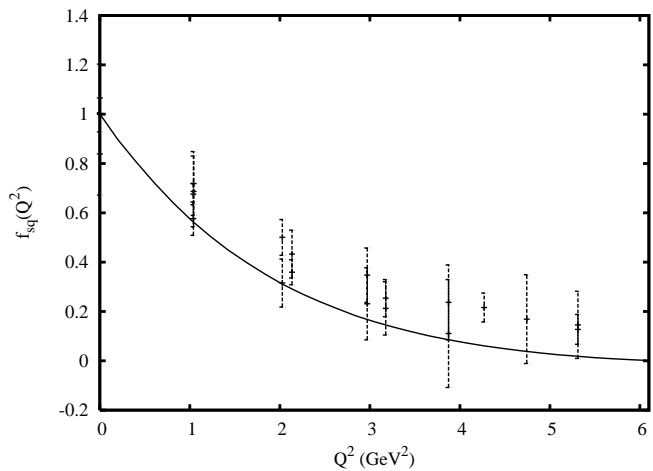}
\caption{The $\chi_{c0}$ Single Quark Form-factor $f_{sq}(Q^2)$.}
\label{chic0Fig}
\end{figure}

A similar procedure can be followed for the vector, scalar, and axial elastic single
quark form factors. The necessary Lorentz decompositions and expressions for the form factors
are given in Appendix \ref{ffApp}. 
The single quark $\chi_{c0}$ elastic form factor for the relativistic BGS+log case
is shown in Fig. \ref{chic0Fig}. The BGS model yields a very similar result and is not shown.
This appears to be generally true and hence most subsequent figures will only display BGS+log
results. As can be seen, the agreement with the lattice data, although somewhat noisy, is very good.

\begin{figure}[h]
\includegraphics[angle=-90,width=8cm]{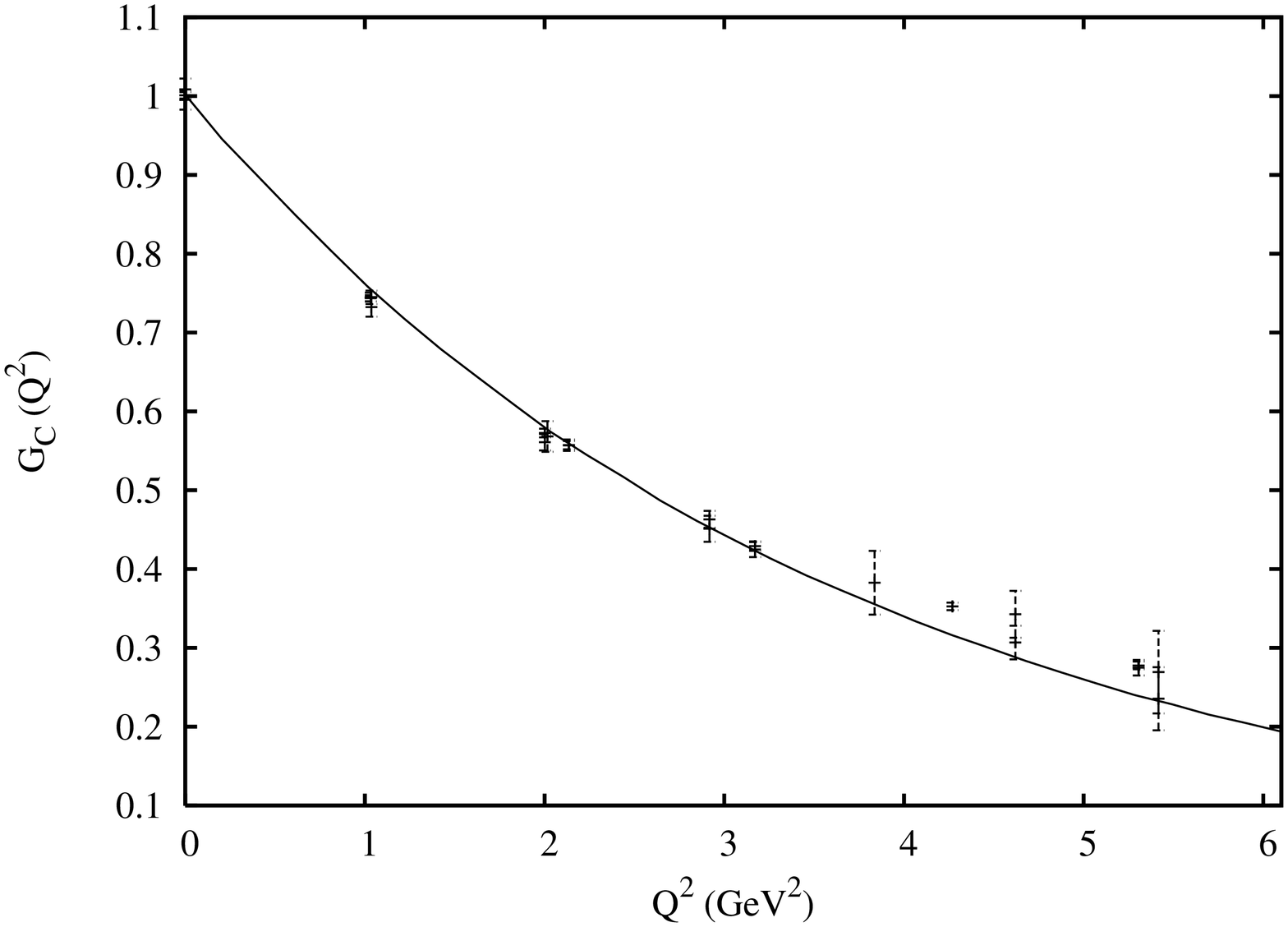}
\includegraphics[angle=-90,width=8cm]{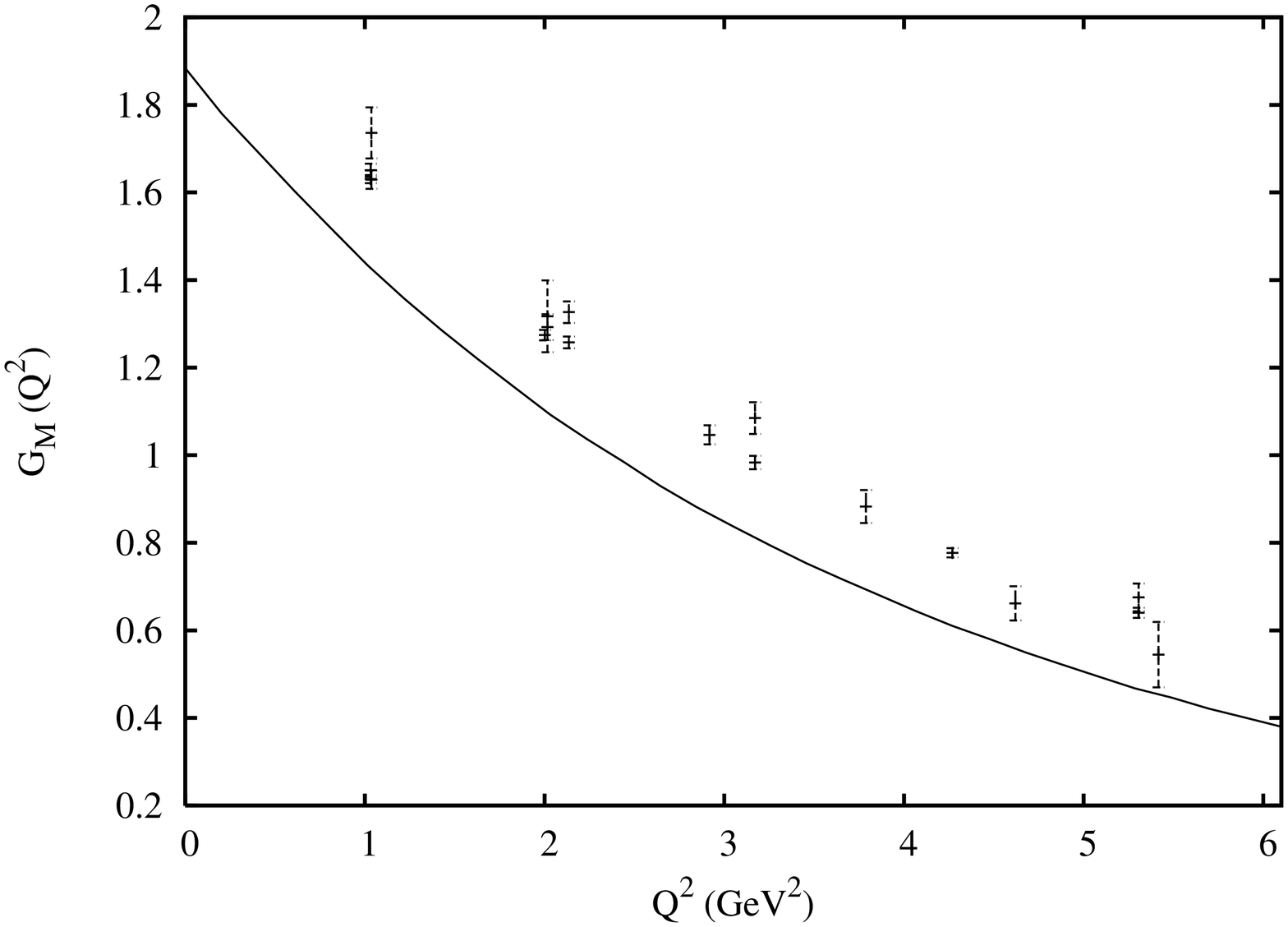}
\caption{Single Quark $J/\psi$ Form Factors $G_{sq}^C$ (left) and $G_{sq}^M$ (right).}
\label{psiFig}
\end{figure}

The left panel of Fig. \ref{psiFig} shows the single 
quark $J/\psi$ charge form factor. The agreement of the relativistic BGS+log
model with the lattice data is remarkable. The right panel of Fig. \ref{psiFig} contains
the magnetic dipole form factor (see Appendix \ref{ffApp} for definitions). In this case 
the form factor at zero recoil is model-dependent.
In the nonrelativistic limit, Eq. \ref{GMEq} implies that $G_M(\vec q = 0) = M_V/m \approx 2$. 
The model prediction is approximately  10\% too small compared to the lattice data. The lattice
results have not been tuned to the physical charmonium masses (charmonium masses are approximately
180 MeV too low); however it is unlikely that this is the source of the discrepancy since the ratio
$M/m$ is roughly constant when $M$ is near the physical mass. Thus it appears that the 
problem lies in the quark model. Reducing the quark mass provides a simple way to improve the 
agreement; however the modifications to the spectrum due to a 10\% reduction in the quark mass 
are difficult to overcome with other parameters while maintaining the excellent agreement with
experiment.

\begin{figure}[h]
\includegraphics[angle=-90,width=8cm]{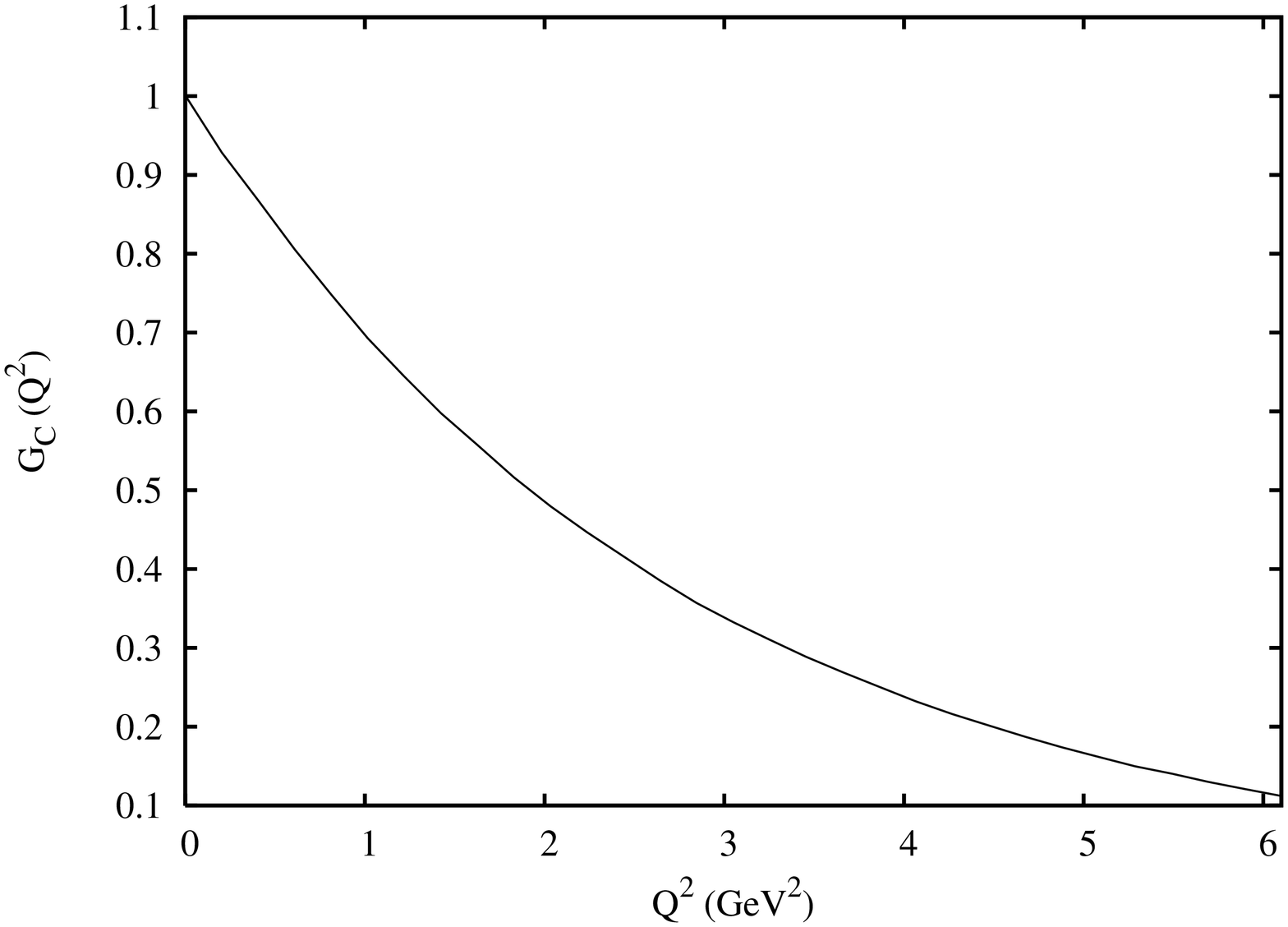}
\includegraphics[angle=-90,width=8cm]{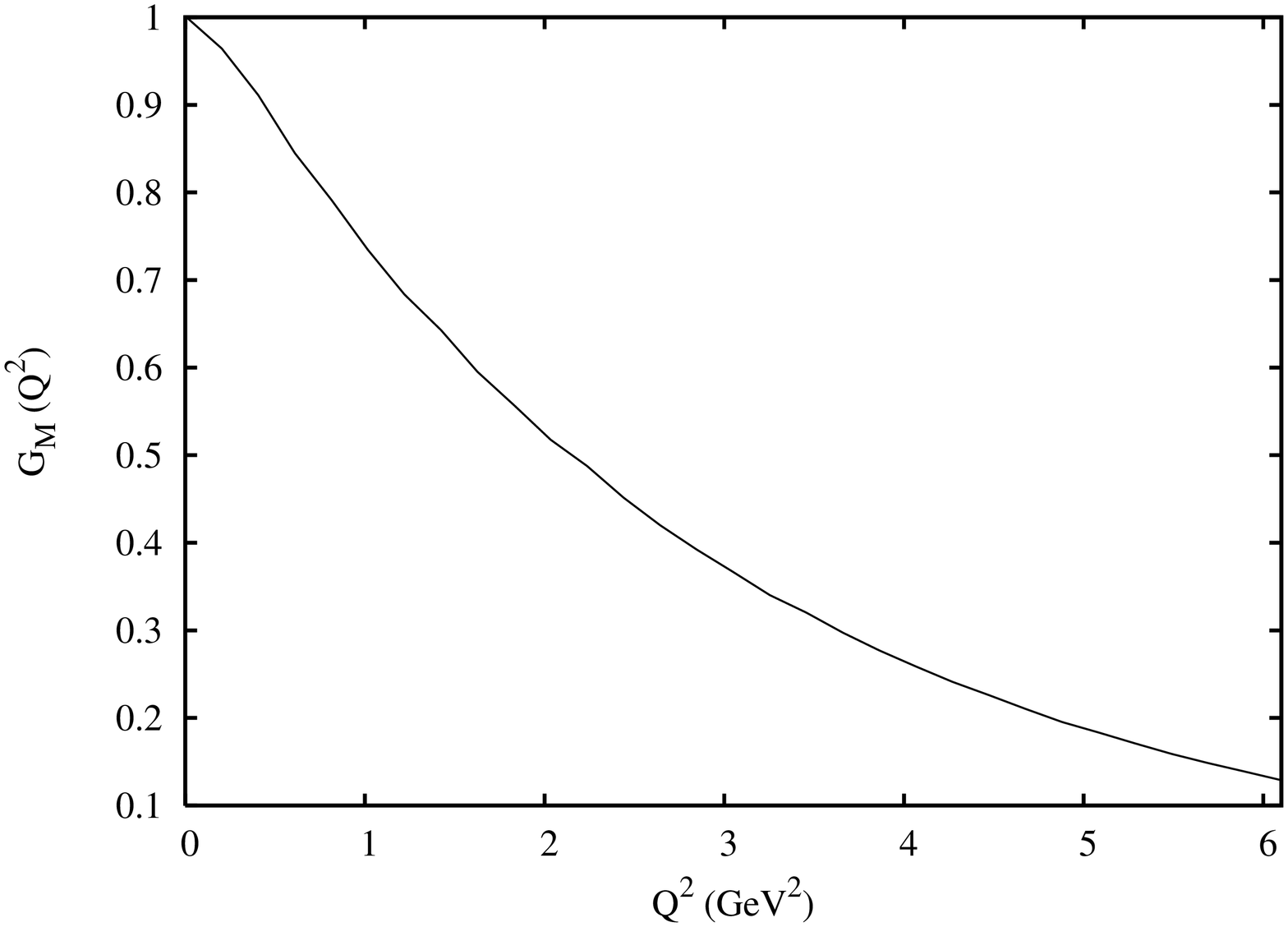}
\caption{Single Quark $h_c$ Form Factors $G_{sq}^C$ (left) and $G_{sq}^M$ (right).}
\label{SqhcFig}
\end{figure}

\begin{figure}[h]
\includegraphics[angle=-90,width=8cm]{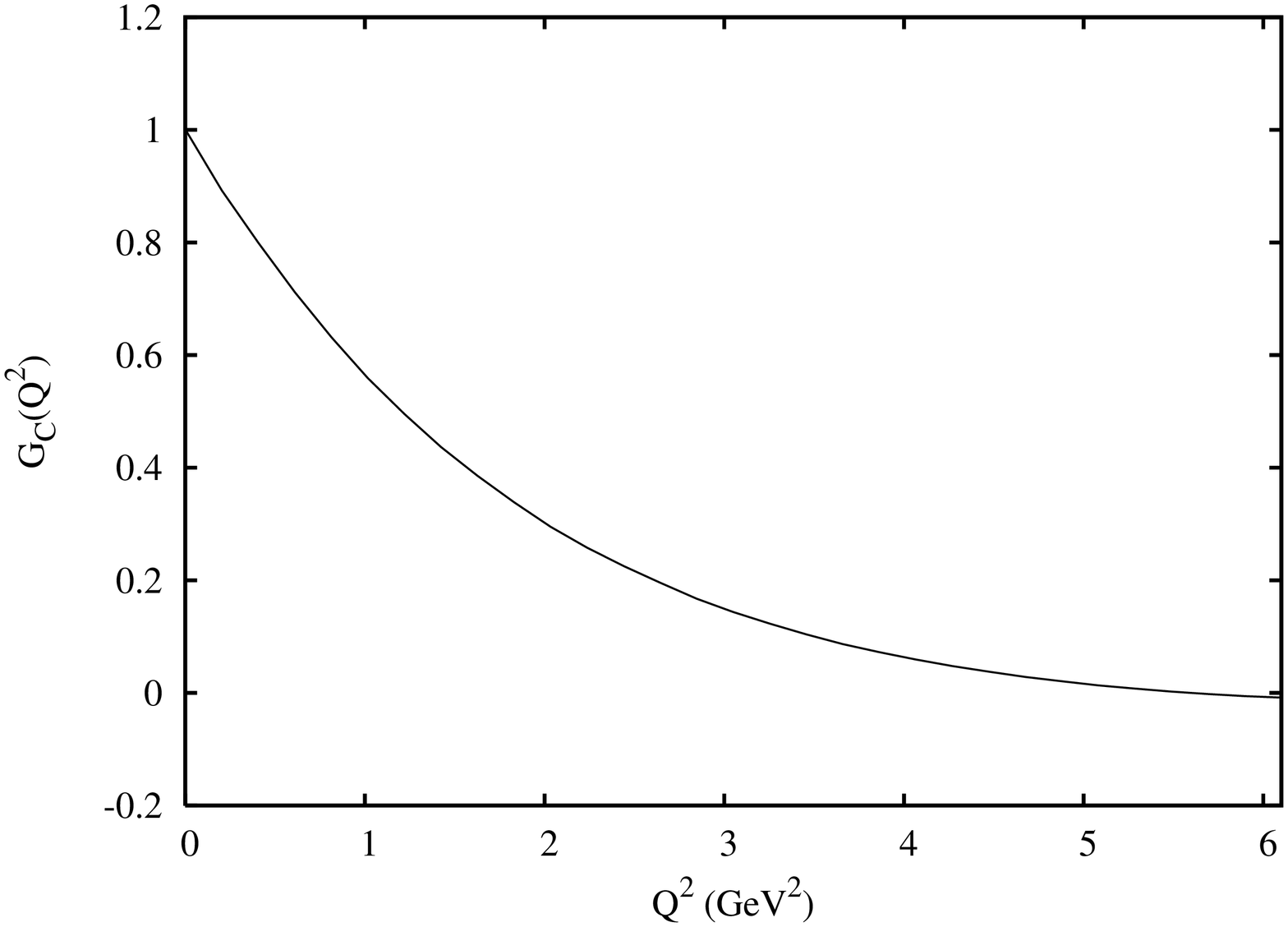} \qquad
\includegraphics[angle=-90,width=8cm]{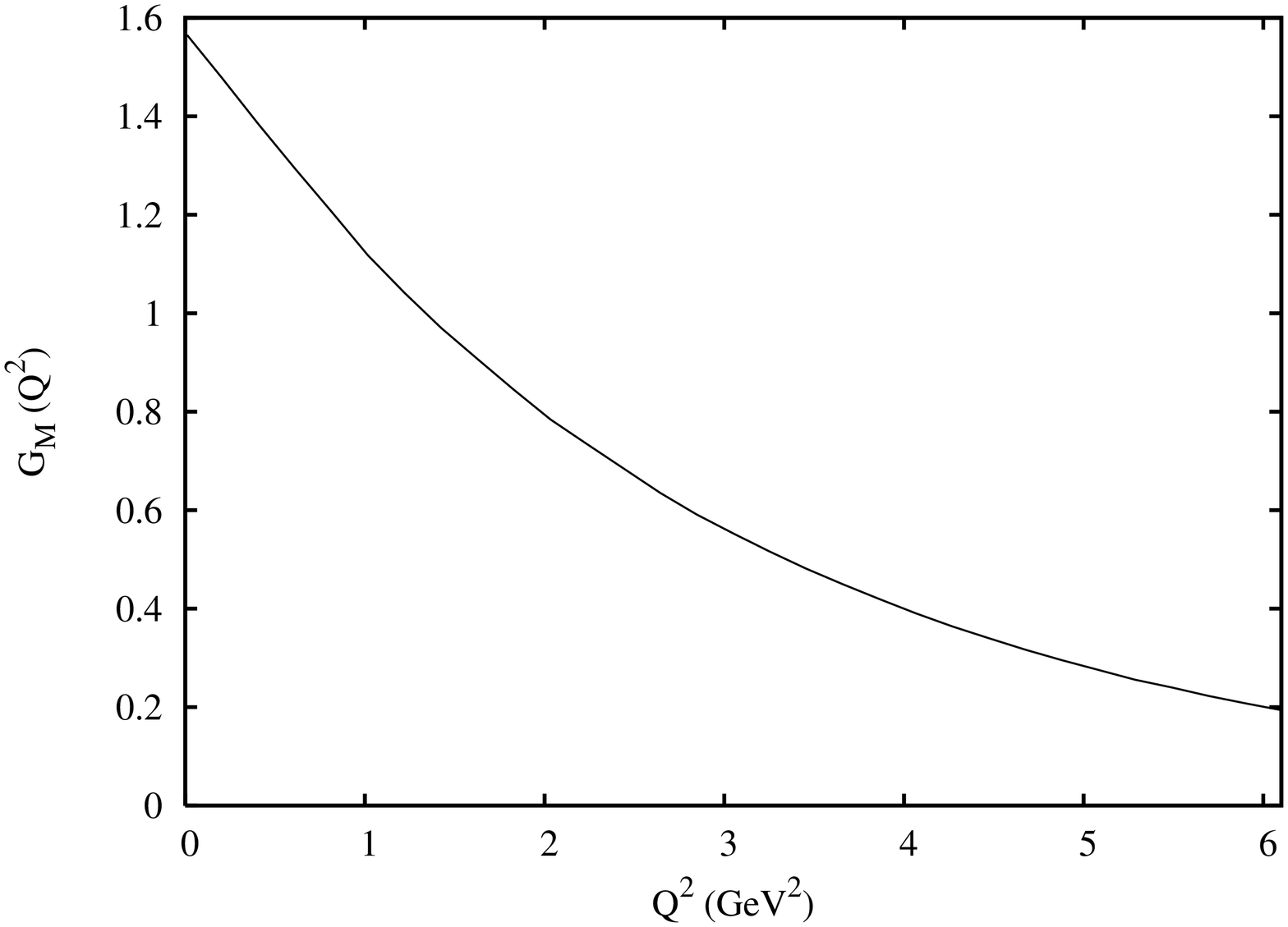}
\caption{Single Quark $\chi_{c1}$ Form Factors $G^C_{sq}$ (left) and $G^M_{sq}$ (right).} 
\label{chic1SQFig}
\end{figure}

Predictions for the single quark elastic electromagnetic form factors of the $h_c$ and $\chi_{c1}$ states
are shown in Figs. \ref{SqhcFig} and \ref{chic1SQFig}. As for the $J/\psi$, the charge form factors are
normalised at zero recoil, while the magnetic form factors take on model-dependent values at zero recoil.
In the nonrelativistic limit these are $G^M_{sq}(\vec q =0) = M/(2m)$ for the $h_c$ and $G^M_{sq}(\vec q = 0) = 3M/(4m)$ for the $\chi_{c1}$.

The presence of a kinematical variable in form factors makes them more sensitive to covariance ambiguities 
than static properties such as decay constants.
In addition to  frame and current component dependence, one also must deal with wavefunction boost
effects that become more pronounced as the recoil momentum increases.
Presumably it
is preferable to employ a frame which minimises wavefunction boost effects since these are not
implemented in the nonrelativistic constituent quark model. Possible choices are (i) the initial meson rest frame
(ii) the final meson rest frame (iii) the Breit frame. These frames correspond to different
mappings of the three momentum to the four momentum: $|\vec q|^2 = Q^2 (1 + \alpha)$ where
$\alpha = 0$ in the Breit frame and $\alpha = Q^2/4 M^2$ in the initial or final rest frame
(these expressions are for elastic form factors with a meson of mass $M$). 
Furthermore, as with decay constants, it is possible to compute the form factors by using different
components of the current.

We consider the $\eta_c$ elastic single quark form factor in greater detail as an example. The 
form factor obtained
from the temporal component of the current in the initial meson rest frame is given in Eqs. \ref{ffFullEq}
and \ref{ff0Eq}. Computing with the spatial components yields Eq. \ref{chiSpatialEq} with the 
nonrelativistic limit

\be
f_{sq}(Q^2) = \frac{2M}{m} \int {d^3 k \over (2\pi)^3} \Phi(\vec{k})\Phi^*\left(\vec{k}+\frac{\vec{q}}{2}\right)\left(\vec k + \frac{\vec q}{2}\right)\cdot \frac{\vec q}{q^2}
\ee
This can be shown to be equivalent to 

\be
\frac{2M}{m} \frac{1}{4}\int d^3x |\Phi(x)|^2 {\rm e}^{-i \vec q \cdot \vec x/2},
\ee
which is Eq. \ref{ff0Eq} in the weak coupling limit. At zero recoil this evaluates to $\frac{M}{2m}$, which is approximately 10\% too small with respect to unity.  Once again, reducing the quark mass
presumably helps improve agreement.

\begin{figure}[h]
\includegraphics[angle=-90,width=8cm]{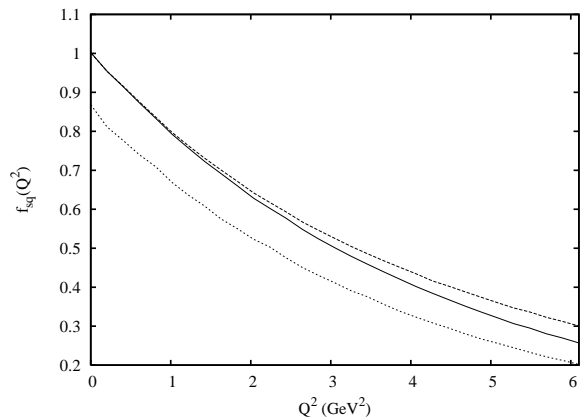}
\caption{Covariance Tests for the Single Quark $\eta_c$ Form Factor.} 
\label{covTestFig}
\end{figure}

Fig. \ref{covTestFig} compares the various methods of computing the $\eta_c$ single quark form factor.
The solid line is the result of Fig. \ref{etaFig}, computed in the initial
rest frame with the temporal component of the current. The dashed line is the computation of the
form factor in the Breit frame. The good agreement is due to a cancellation between the different
four-vector mapping discussed above and the modifications induced by computing the quark model form factor
in the Breit frame. The lower dashed line is the form factor computed from the spatial components
of the current (Eq. \ref{chiSpatialEq}). It is evidently too small compared to the correctly normalised
results by approximately a factor
of $2m/M$, indicating that the method is accurate at the 10\% level.

Finally, the large $Q^2$ behaviour of pseudoscalar form factors is a controversial topic. We do not
presume to resolve the issues here; rather we note that the preferred method for obtaining the
form factor yields an asymptotic behaviour proportional to  $\alpha_s(Q^2) f_{Ps} M_{Ps} /Q^2$, 
which is similar,
 but not identical, to that expected in perturbative QCD\cite{FJ}.
Nevertheless, the model is not applicable in this regime and the asymptotic scaling should not be
taken seriously.

%

\section{Charmonium Transition Form Factors} 
\label{tffSect}

Transition form factors 
convolve differing wavefunctions and therefore complement the information contained in
single quark elastic form factors.
They also have the important benefit of being experimental observables at $Q^2 = 0$.

The computation of transition form factors proceeds as for elastic form factors, with the exception
that the current is coupled to all quarks. Lorentz decompositions and
quark model expressions for a variety of transitions are presented in App. \ref{ffApp}.  The mapping
between three-momentum and $Q^2$ is slightly different in the case of transition form factors. In
the Breit frame this is
\be
|\vec q|^2 = Q^2 + {(m_2^2 - m_1^2)^2 \over Q^2 + 2m_1^2 + 2 m_2^2},
\ee
while in the initial rest frame it is
\be
|\vec q|^2 = {Q^4 + 2Q^2 (m_1^2+m_2^2) + (m_1^2 - m_2^2)^2\over 4 m_1^2}.
\label{generalQsqEq}
\ee
An analogous result holds for the final rest frame mapping.

\begin{figure}[h]
\includegraphics[angle=-90,width=8cm]{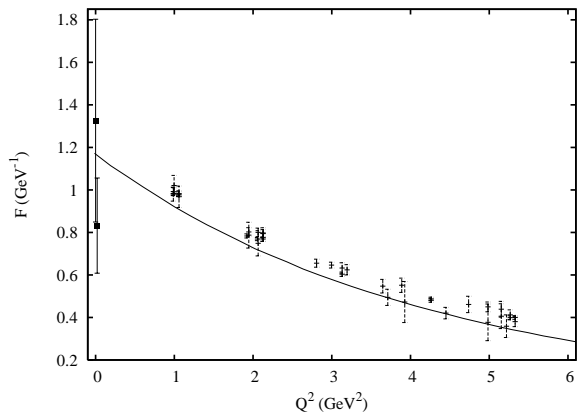}
\caption{Form Factor $F(Q^2)$ for $J/\psi\rightarrow \eta_c\gamma$. Experimental points are indicated with squares.}
\label{FpsiFig}
\end{figure}

\begin{figure}[h]
\includegraphics[angle=-90,width=8cm]{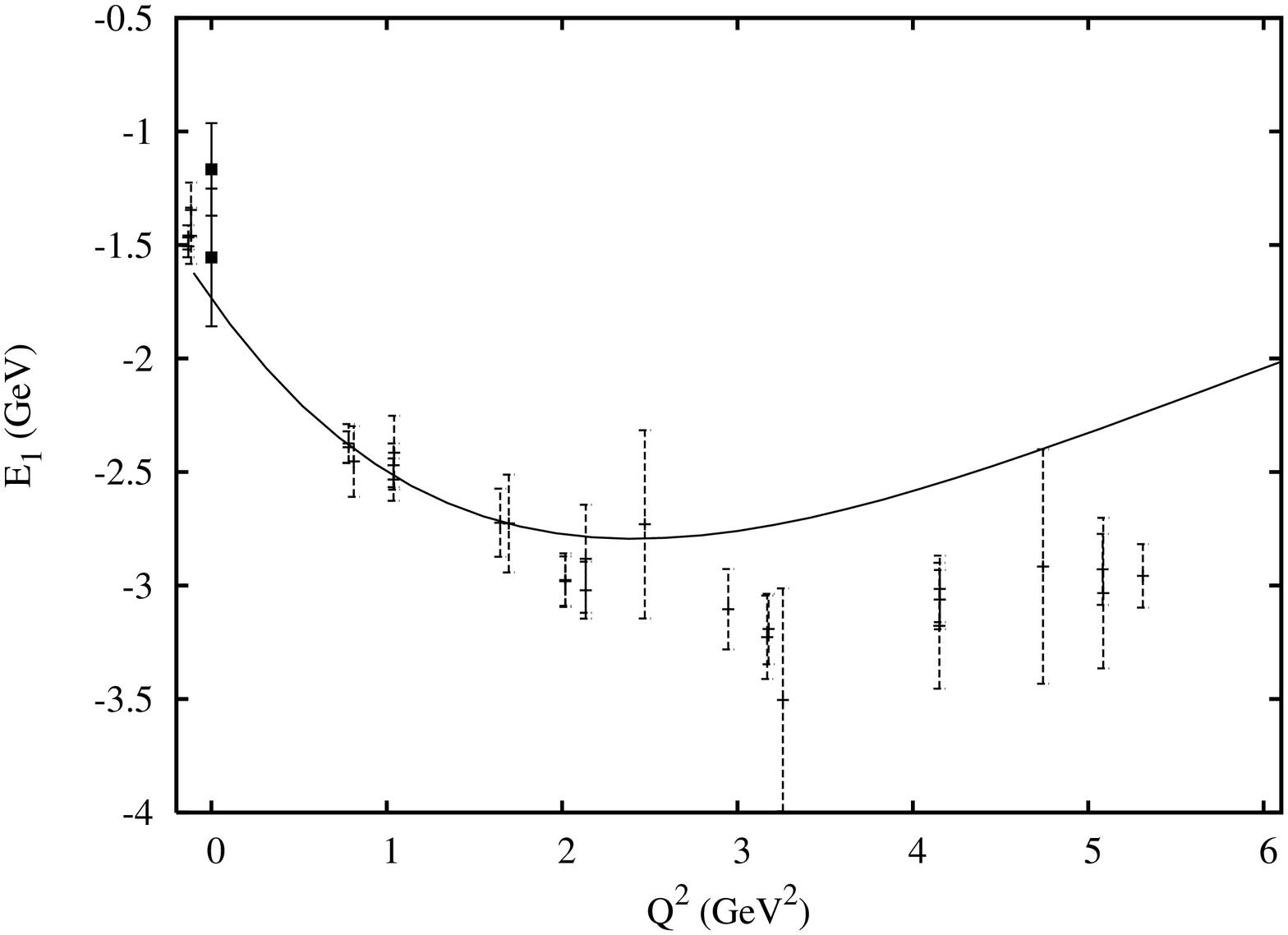} \qquad
\includegraphics[angle=-90,width=8cm]{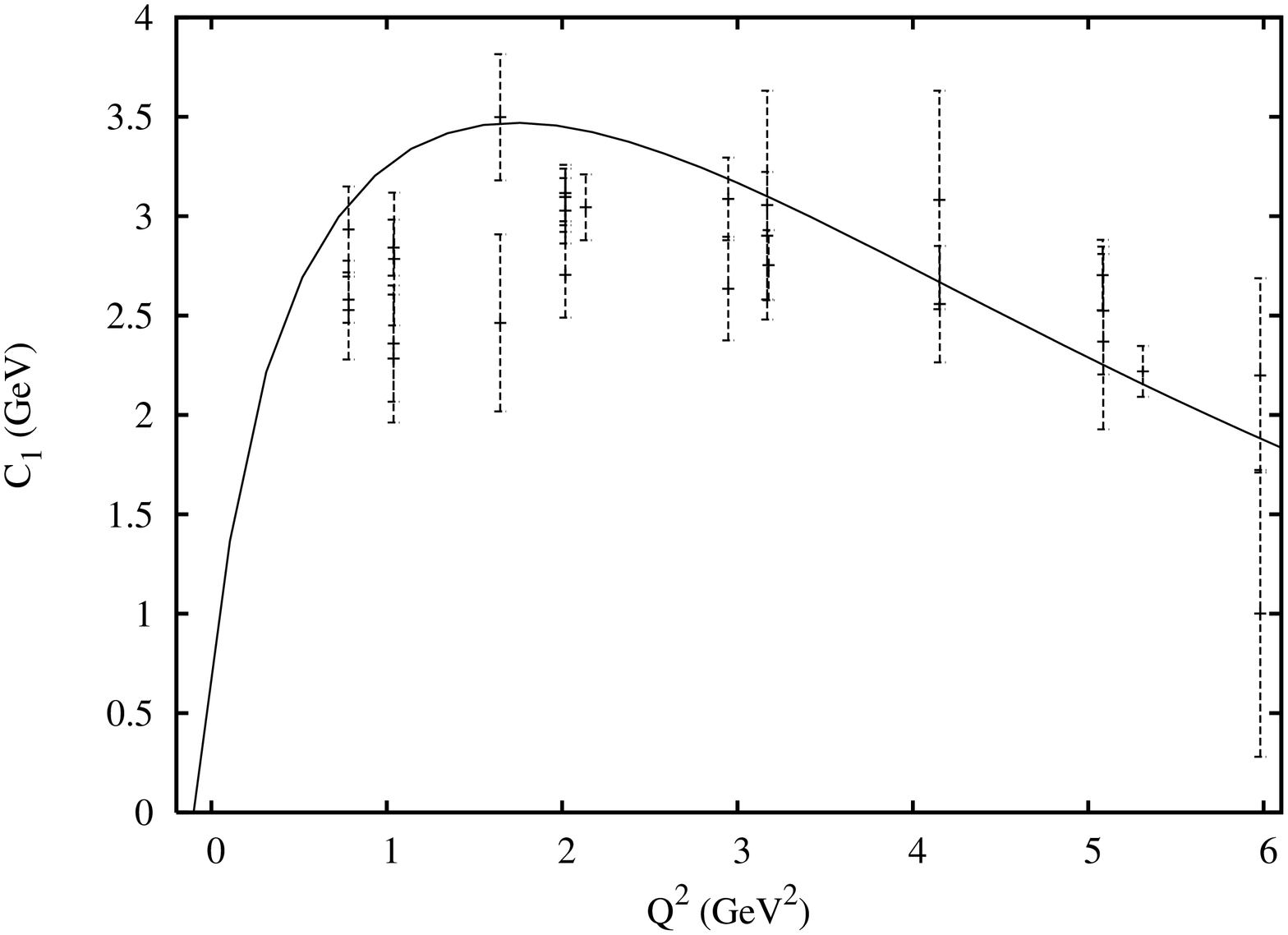}
\caption{Form Factors $E_1(Q^2)$ (left) and $C_1(Q^2)$ (right) for $\chi_{c0}\rightarrow J/\psi\gamma$.
Experimental points are indicated with squares.}
\label{E1Fig}
\end{figure}

\begin{figure}[h]
\includegraphics[angle=-90,width=8cm]{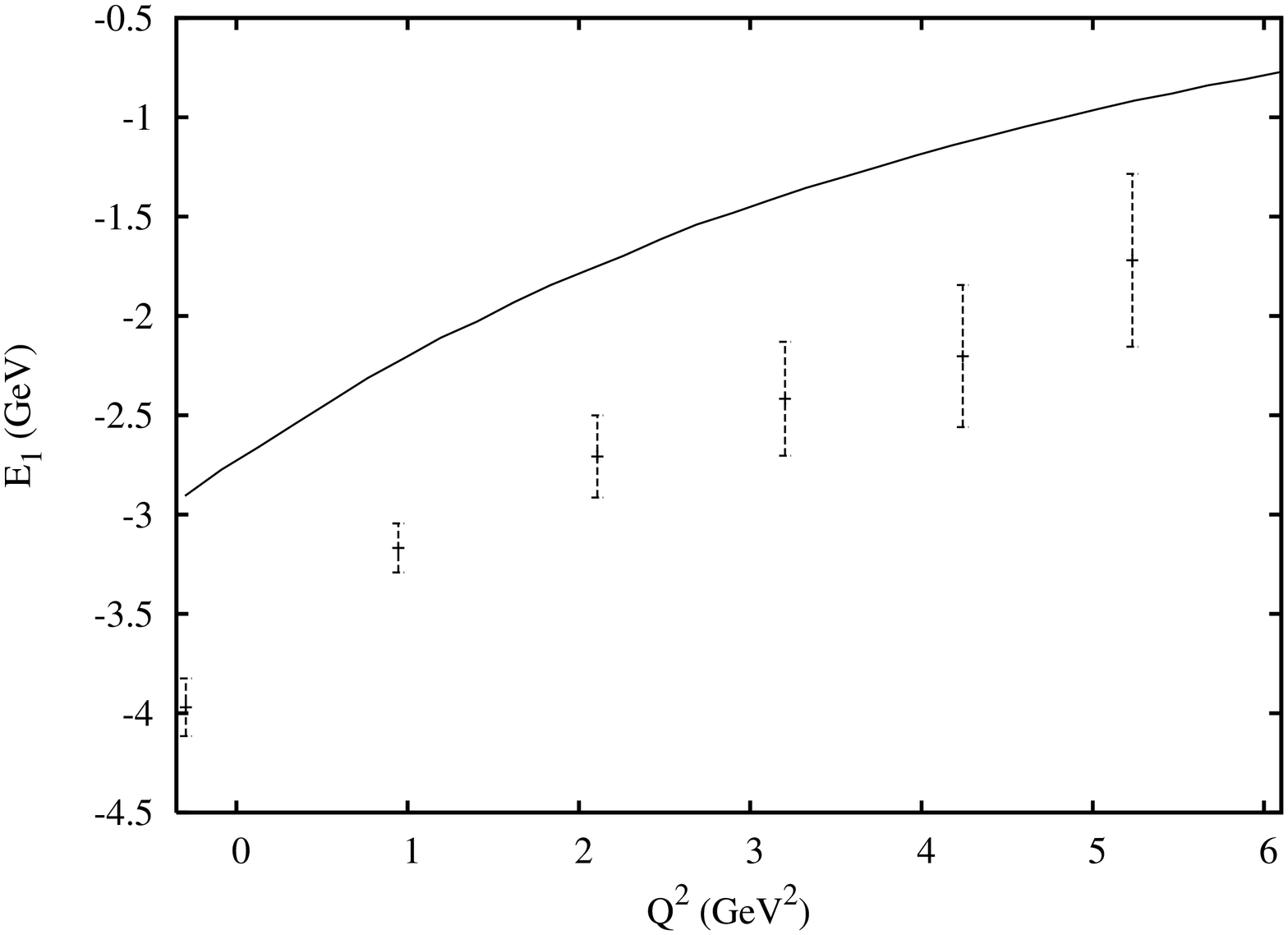} \qquad
\includegraphics[angle=-90,width=8cm]{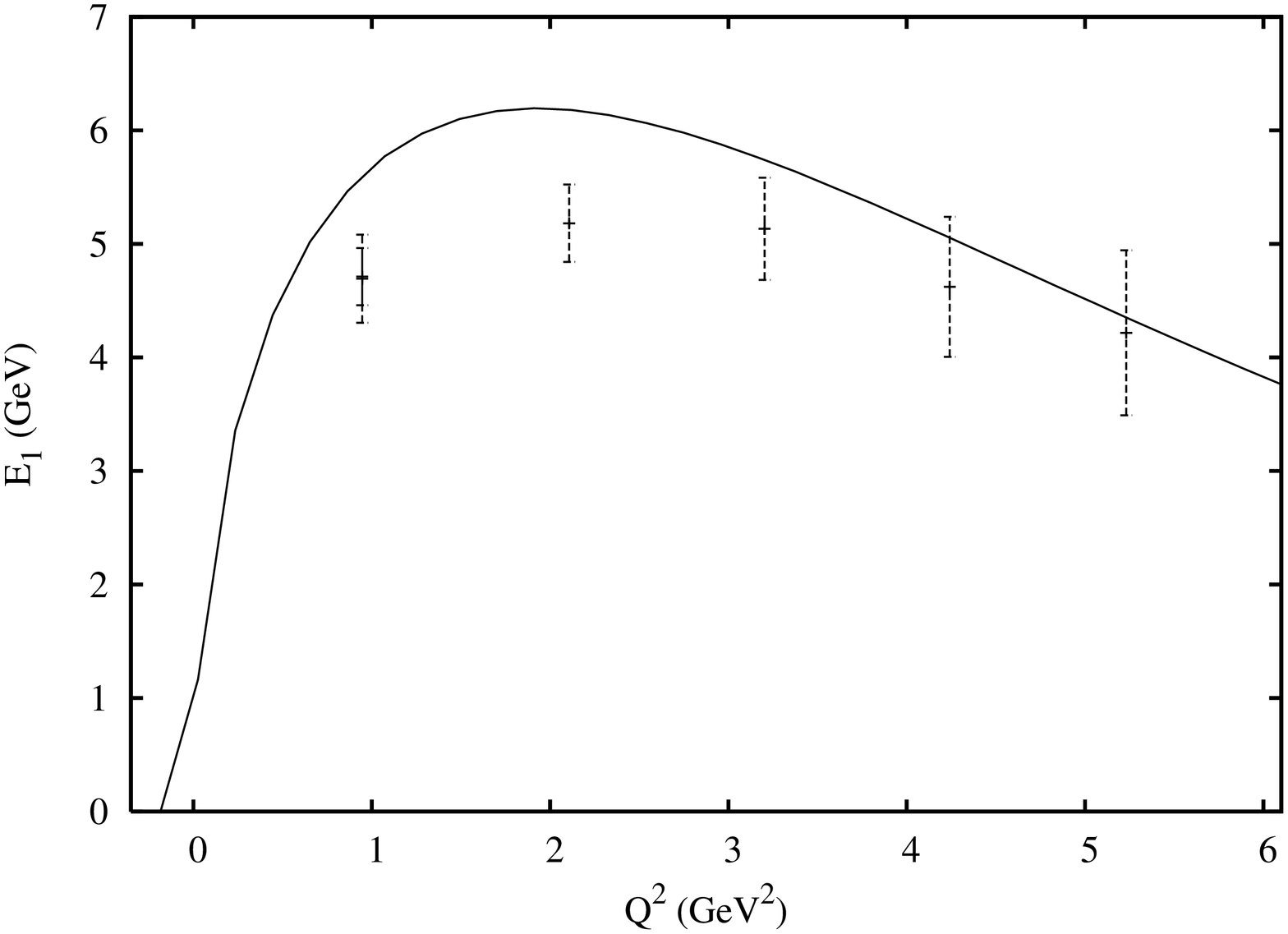} 
\caption{Form Factors  $E_1(Q^2)$ (left) and $C_1(Q^2)$ (right) for $h_c\rightarrow \eta_c\gamma$.}
\label{hcFig}
\end{figure}

\begin{figure}[h]
\includegraphics[angle=-90,width=8cm]{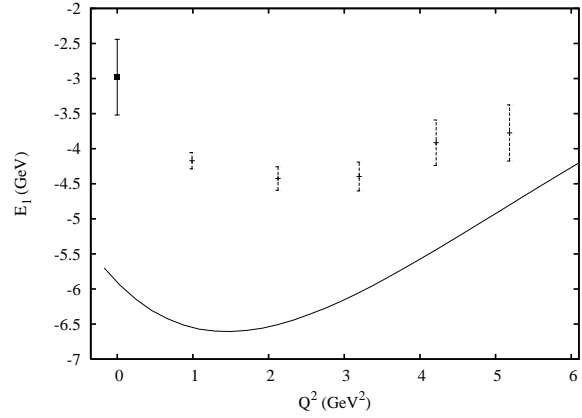}
\caption{Form Factor $E_1(Q^2)$ for $\chi_{c1} \to J/\psi\gamma$.  Experimental points are indicated with squares.}
\label{chi1psiE1Fig}
\end{figure}

\begin{figure}[h]
\includegraphics[angle=-90,width=8cm]{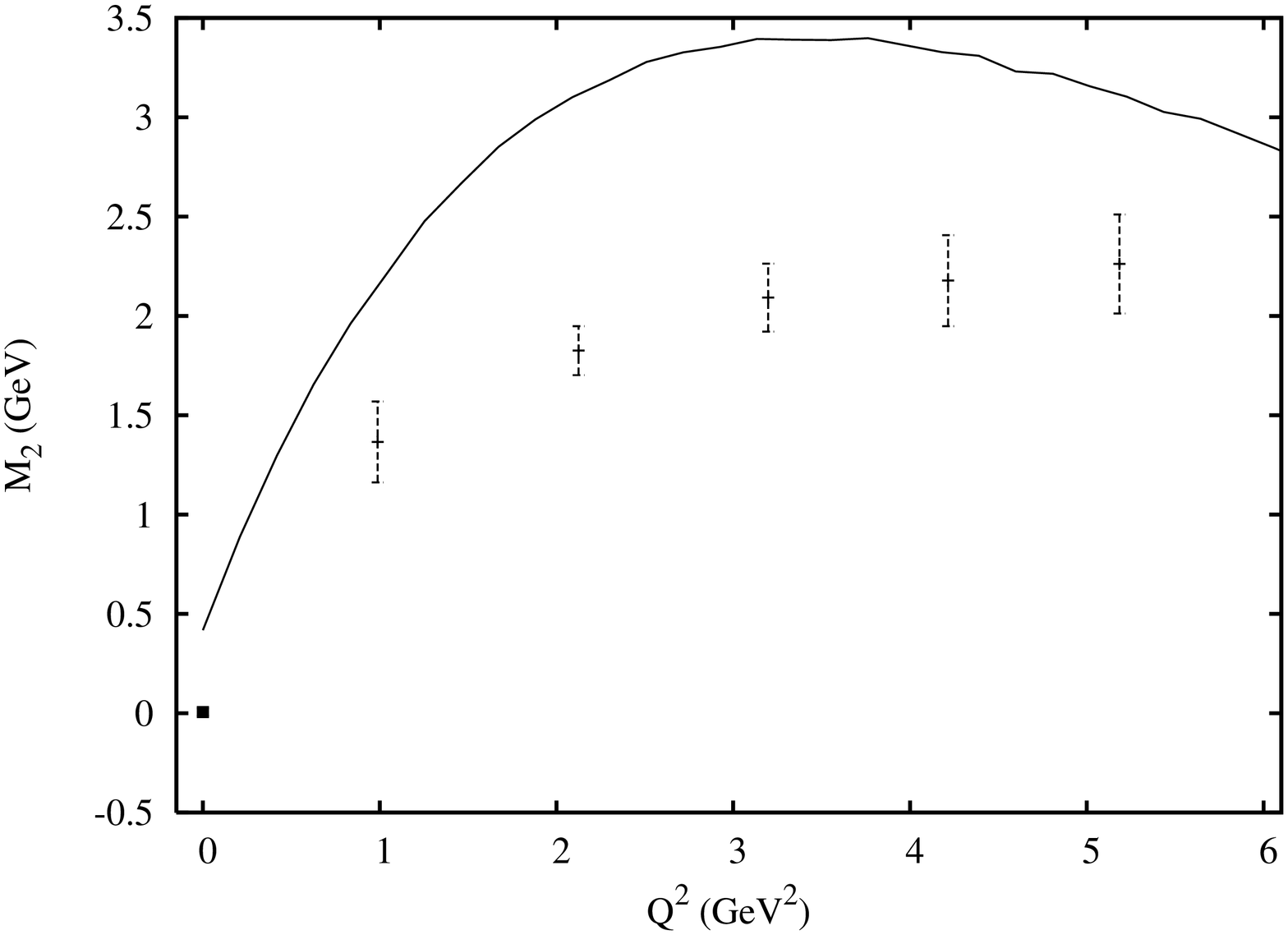} \qquad
\includegraphics[angle=-90,width=8cm]{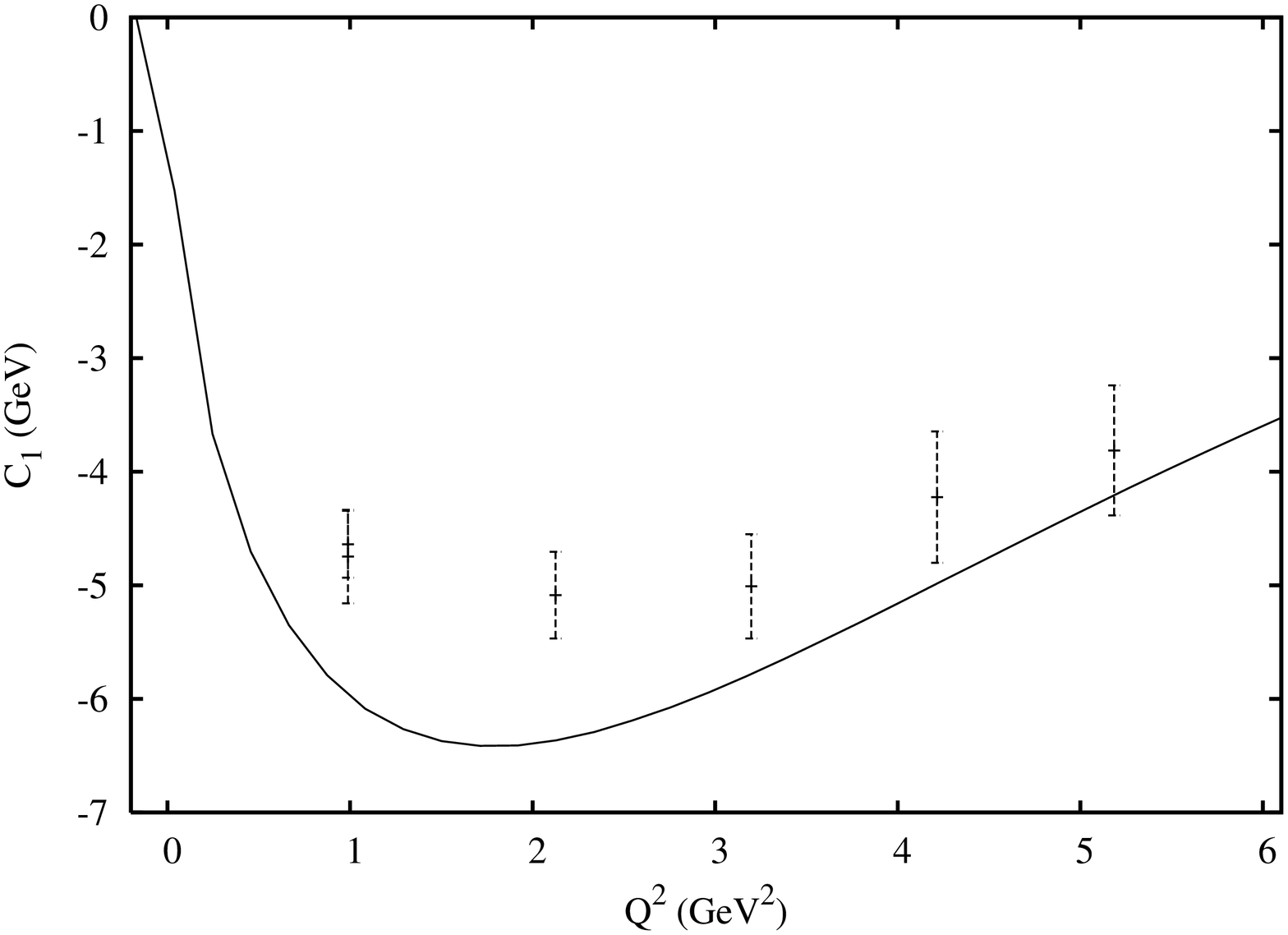}
\caption{Form Factors $M_2(Q^2)$ (left) and $C_1(Q^2)$ (right) for $\chi_{c1}\rightarrow J/\psi\gamma$. Experimental points are indicated with squares.}
\label{chi1psiM2Fig}
\end{figure}

Computed form factors are compared to the lattice calculations of Ref. \cite{JLlatt} and experiment (where
available) in Figs. \ref{FpsiFig} to \ref{chi1psiM2Fig}. 
Experimental measurements (denoted by squares in the figures) have been determined as follows:
For $J/\psi \to \eta_c\gamma$ Crystal Barrel\cite{CB} measure $\Gamma = 1.14 \pm 0.33$ keV. Another
estimate of this rate may be obtained by combining the  Belle measurement\cite{belle} of $\Gamma(\eta_c \to \phi\phi)$ with the rate for $J/\psi \to \eta_c\gamma \to \phi\phi\gamma$ reported in the PDG\cite{PDG}.
One obtains $\Gamma(J/\psi \to \eta_c\gamma) = 2.9 \pm 1.5$ keV\cite{JLlatt}. Both these data are displayed in
Fig. \ref{FpsiFig}.

Two experimental points for $\chi_{c0} \to J/\psi\gamma$ are displayed in Fig. \ref{E1Fig} (left panel). These
correspond to the PDG value $\Gamma(\chi_{c0} \to J/\psi \gamma) = 115 \pm 14$ keV and a recent result from
CLEO\cite{cleo}: $\Gamma(\chi_{c0}\to J/\psi \gamma) = 204 \pm 31$ keV.

Finally, the experimental points for the $E_1$ and $M_2$ $\chi_{c1} \to J/\psi \gamma$ multipoles (Fig. 
\ref{chi1psiM2Fig}) are determined from the decay rate reported in the PDG and the ratio 
$M_2/E_1 = 0.002 \pm 0.032$ determined by E835\cite{E835}.

Overall the agreement between the model, lattice, and experiment is impressive.  The exception is the $E_1$
multipole for $\chi_{c1} \to J/\psi \gamma$. We have no explanation for this discrepancy. Note that the
quenched lattice and quark model both neglect coupling to higher Fock states, which could affect the observables.
The agreement with experiment indicates that such effects are small (or can be effectively subsumed into quark 
model parameters and the lattice scale), thereby justifying the use of the quenched approximation and the 
simple valence quark model when applied to these observables.

\begin{figure}[h]
\includegraphics[angle=-90,width=8cm]{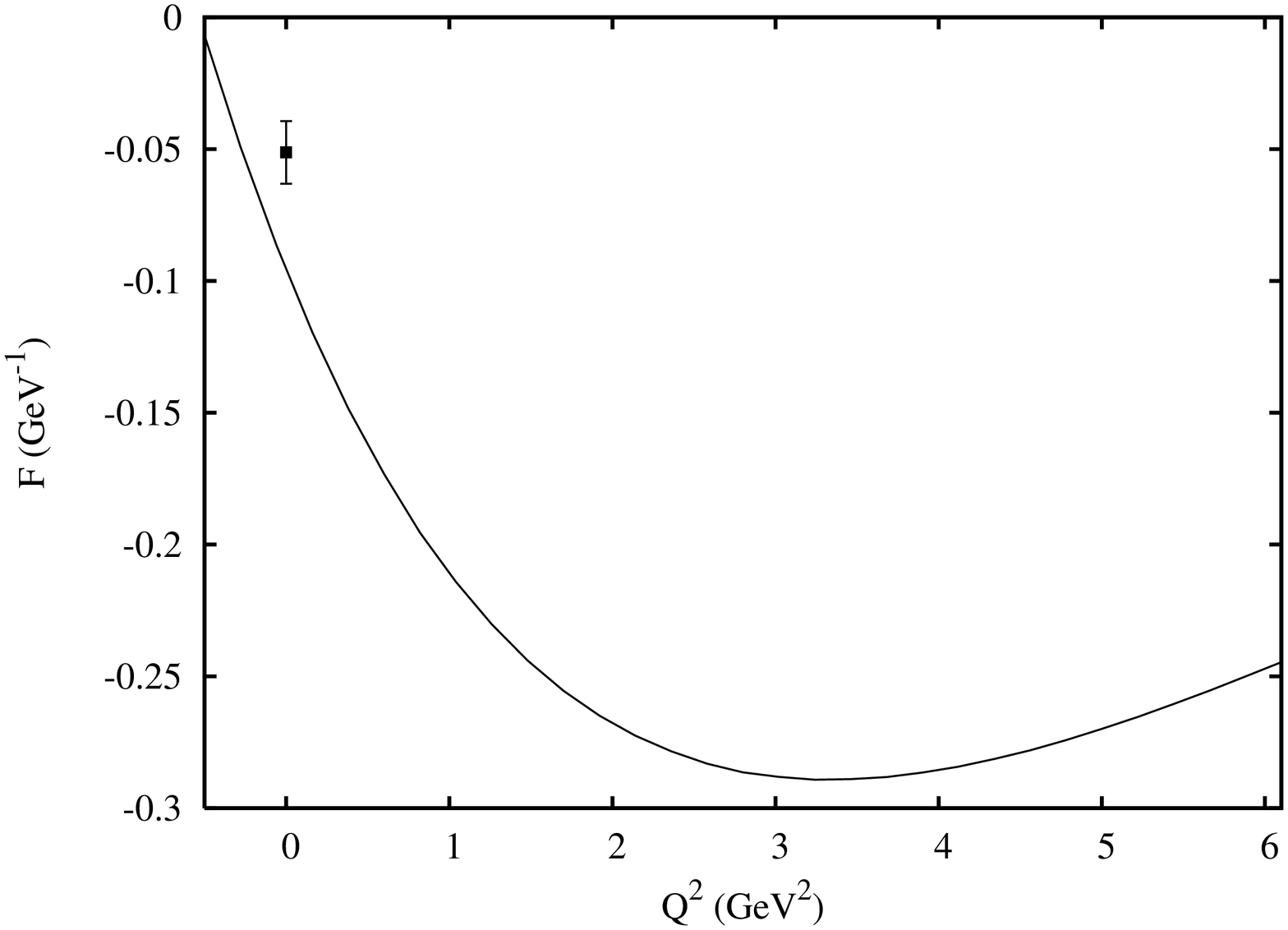} \qquad
\includegraphics[angle=-90,width=8cm]{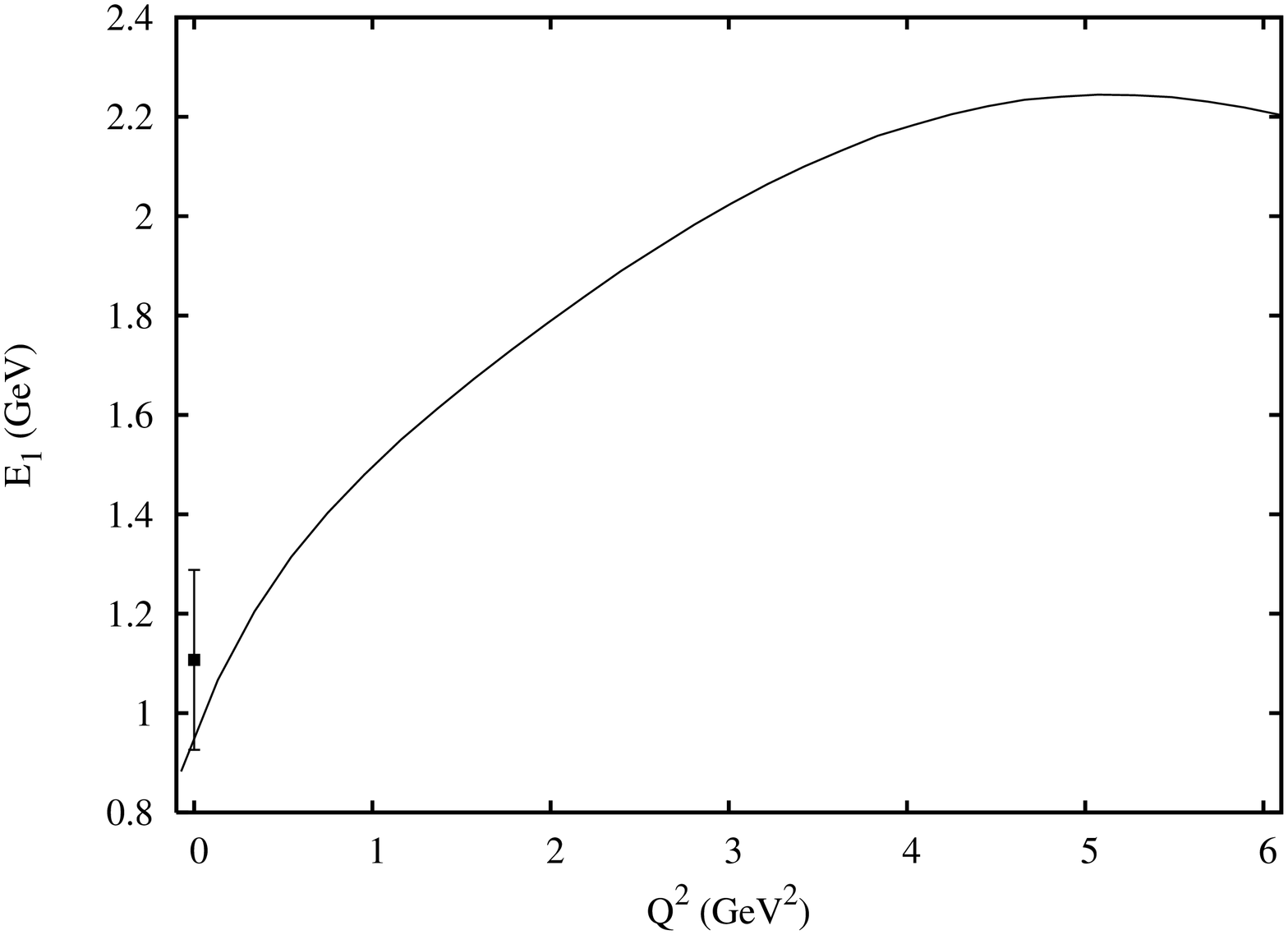}
\caption{Form Factor $F(Q^2)$ for $\psi' \to \eta_c \gamma$ (left). Form factor $E_1(Q^2)$ for $\psi' \to \chi_{c0}\gamma$ (right). Experimental points are indicated with squares.}
\label{psi2etaFig}
\end{figure}

Predictions for excited state form factors are simple to obtain in the quark model (in contrast to
lattice gauge theory, where isolating excited states is computationally difficult). Two examples are
presented in Fig. \ref{psi2etaFig}. The agreement with experiment (squares) is acceptable.

\section{Charmonium $\gamma\gamma$ Widths}  
\label{ggSect}

Two-photon decays of mesons are of considerable interest as a search mode, a probe
of internal structure, and as a test of nonperturbative QCD modelling. An illustration
of the importance of the latter point is the 
recent realization that the usual factorisation
approach to orthopositronium (and its extensions to QCD) decay violates low energy 
theorems\cite{ps}.

\subsection{Formalism and Motivation}

It has been traditional to compute decays such as $Ps \to \gamma\gamma$ by assuming 
factorisation between soft bound state dynamics and hard rescattering into photons\cite{pw}.
This approximation is valid when the photon energy is much greater than the binding
energy $E_B \sim m\alpha^2$. This is a difficult condition to satisfy in the case
of QCD where $\alpha \to \alpha_s \sim 1$. Nevertheless, this approach has been adopted
to inclusive strong decays of mesons\cite{ap,brodsky,kwong} and has been extensively applied to
two-photon decays of quarkonia\cite{bc}. 

The application of naive factorisation to orthopositronium decay (or
$M \to ggg$, $\gamma gg$ in QCD) leads to a differential decay rate that scales as $E_\gamma$ for small photon
energies\cite{op} -- at odds with the $E_\gamma^3$ behaviour required by gauge invariance and
analyticity (this is Low's theorem\cite{low}). The contradiction can be traced to the scale dependence
of the choice of relevant states and can be resolved with a careful NRQED analysis\cite{mr}. For example,
a parapositronium-photon intermediate state can be important in orthopositronium decay at low energy.
Other attempts to address the problem by treating binding energy nonperturbatively can be found
in Refs. \cite{smith,ni}.

\begin{figure}[h]
\includegraphics[angle=0,width=5cm]{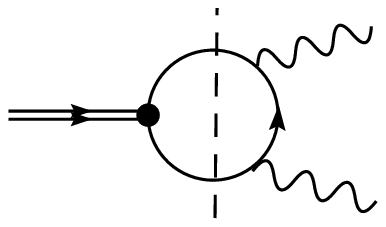}
\caption{Naive Factorisation in Positronium Decay.}
\label{posDecayFig}
\end{figure}

Naive factorisation is equivalent to making a vertical cut through the loop diagram representing
$Ps \to n\gamma$\cite{smith} (see Fig. \ref{posDecayFig}). Of course this ignores cuts across photon vertices that correspond to
the neglected intermediate states mentioned above. In view of this, a possible improvement is
to assume that pseudoscalar meson decay to two photons occurs via an intermediate vector meson
followed by a vector meson dominance transition to a photon. This approach was indeed suggested
long ago by van Royen and Weisskopf\cite{vRW} who made simple estimates of the rates for $\pi^0\to \gamma\gamma$ and $\eta \to \gamma\gamma$. This proposal is also in accord with time ordered perturbation 
theory applied to QCD in Coulomb gauge, where intermediate bound states created by instantaneous gluon
exchange must be summed over.

Finally, 
one expects that an effective description should work for sufficiently low momentum
photons. The effective Lagrangian for pseudoscalar decay can be written as 

\be
{\cal L} = g\int \eta F^{\mu\nu}\tilde F_{\mu\nu}
\ee
leading to the prediction $\Gamma(\eta\to \gamma\gamma) \propto g^2 m_\eta^3$. 
Since this scaling with respect to the pseudoscalar mass appears to be experimentally satisfied for
$\pi$, $\eta$, $\eta'$ mesons, Isgur {\it et al.} inserted an {\it ad hoc} dependence of
$m_\eta^3$ in their quark model computations\cite{HI,GI}.
While perhaps of practical use, this
approach is not theoretically justified and calls into doubt the utility of the quark model in
this context. Indeed simple quark model computations of the amplitude of Fig. \ref{posDecayFig}
are not dependent on binding energies and can only depend on kinematic quantities such as quark
masses.

In view of the discussion above, we chose to abandon the factorisation approach and compute two-photon 
charmonium decays in the quark model
in bound state time ordered perturbation theory. This has the effect of saturating the intermediate state
with all possible vectors, thereby bringing in binding energies, a nontrivial dependence
on the pseudoscalar mass, and incorporating oblique cuts in the loop diagram. 


\subsection{Results}

The general amplitude for two-photon decay of pseudoscalar quarkonium can be written as

\be
{\cal A}(\lambda_1 p_1; \lambda_2 p_2) = \epsilon^*_\mu(\lambda_1,p_1) \epsilon^*_\nu(\lambda_2, p_2) {\cal M}^{\mu\nu}
\ee
with
\be
{\cal M}^{\mu\nu}_{Ps} = i M_{Ps}(p_1^2, p_2^2, p_1\cdot p_2) \, \epsilon^{\mu\nu\alpha\beta} \, p_{1\alpha}p_{2\beta}.
\ee
The total decay rate is then $\Gamma(Ps \to \gamma\gamma) = {m_{Ps}^3 \over 64 \pi} |M_{Ps}(0,0)|^2$.

Before moving on to the quark model computation, 
it is instructive to evaluate the amplitude in an effective field theory that incorporates
pseudoscalars, vectors, and vector meson dominance. The relevant Lagrangian
density is

\begin{equation}
{\cal L} = -i Q m_V f_V V_\mu A^\mu - \frac{1}{2} Q F^{(V)} \eta \tilde F_{\mu\nu} V^{\mu\nu}
\end{equation}
where $\tilde F^{\mu\nu} = \frac{1}{2}\epsilon^{\mu\nu\alpha\beta}F_{\alpha\beta}$ and
$V^{\mu\nu} = \partial^\mu V^\nu - \partial^\nu V^\mu$\footnote{The vector meson dominance
term is not gauge invariant. Why this is not relevant here is discussed in Sect. 15 of Ref. \cite{RPF}.}.
Evaluating the transition $Ps \to \gamma\gamma$ yields

\begin{equation}
M_{Ps}(p_1^2,p_2^2) = \sum_V m_V f_V Q^2 \left({F^{(V)}(p_1^2)\over p_2^2-m_V^2} + {F^{(V)}(p_2^2)\over p_1^2-m_V^2}\right).
\label{relMPsEq}
\end{equation}
Hence the pseudoscalar decay rate is 

\begin{equation}
\Gamma(Ps \to \gamma\gamma) = {m_{Ps}^3 Q^4 \over 16 \pi} \left( \sum_V {f_V F^{(V)}(0) \over m_V}\right)^2.
\label{relRateEq}
\end{equation}
Notice that the desired cubic pseudoscalar mass dependence is achieved in a simple manner in this 
approach.

The application of this formula is complicated by well-known ambiguities in the vector meson dominance
model (namely, is $p_V^2 = m_V^2$ or zero?). The time ordered perturbation theory of the quark model
suffers no such ambiguity (although, of course, it is not covariant) and it is expedient
to use the quark model to resolve the ambiguity. We thus choose to evaluate
the form factor at the kinematical point $|\vec q| = m_{Ps}/2$, appropriate to $Ps \to \gamma\gamma$ in
the pseudoscalar rest frame. Applying  Eq. \ref{generalQsqEq} to the virtual process 
$\eta_c \to J/\psi \gamma$ then implies that the argument of the form factor should be $Q^2 = 2.01$ GeV$^2$. 

A simple estimate of the rate for $\eta_c \to \gamma\gamma$ can now be obtained from
Eq. \ref{relRateEq},
$f_{J/\psi} \approx 0.4$ GeV, and $F^{(V)}(Q^2=2\ {\rm GeV}^2) \approx 0.7$ GeV$^{-1}$ 
(Fig. \ref{FpsiFig}). The result is $\Gamma(\eta_c\to \gamma\gamma) \approx 7.1$ keV, in reasonable 
agreement with experiment.

\begin{figure}[h!]
\includegraphics[angle=270,width=8cm]{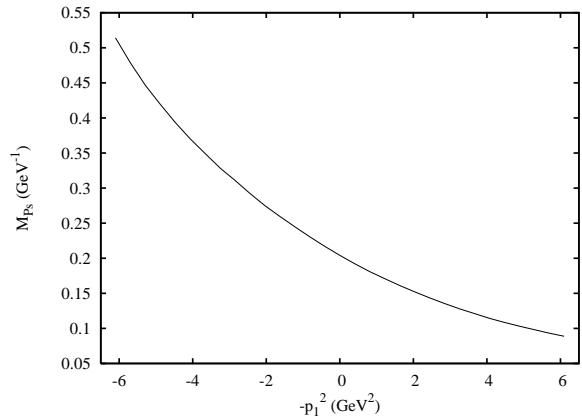}
\caption{The Two-photon Form Factor $M_{Ps}(p_1^2,p_2^2=0)$ for $\eta_c \to \gamma\gamma$.}
\label{MPsFig}
\end{figure}

Finally, the predicted form of the two-photon $\eta_c$ form factor is shown in Fig. \ref{MPsFig} in
the case that one photon is on-shell. The result is a slightly distorted monopole (due to vector
resonances and the background term in Eq. \ref{relMPsEq}) that disagrees strongly with naive
factorisation results. Lattice computations should be able to test this prediction\cite{JD} -- if
it is confirmed, the factorisation model will be strongly refuted.

As motivated above, the microscopic description of the $\eta_c$ two-photon decay is best evaluated
in bound state time ordered perturbation theory. Thus one has

\be
{\cal A}_{NR} = \sum_{\gamma,V}{\langle \gamma(\lambda_1,p_1) \gamma(\lambda_2,p_2) | H | \gamma, V\rangle \, 
\langle \gamma, V | H | Ps \rangle \over (m_{Ps} - E_{\gamma V}) }
\label{microAEq}
\ee
The second possible
time ordering requires an extra vertex to permit the transition $\langle Ps, V | \gamma\rangle$ 
and hence is higher order in the Fock space expansion. Thus the second
time ordering has been neglected in Eq. \ref{microAEq}.

The amplitudes  can be written in terms of the relativistic decompositions of the previous sections.
One obtains the on-shell amplitude

\be
M_{Ps} =
\sum_{V}Q^2 \sqrt{\frac{m_V}{E_V}} f_V  \frac{F^{(V)}(q)}{m_{Ps}-E_{\gamma V}(q)}.
\label{MPsEq}
\ee
We choose to label the momentum dependence with the nonrelativistic $q=|\vec q|$ in these expressions
\footnote{The naive application of the method advocated here to light quarks will fail. 
In this case the axial anomaly requires that 
$M_{Ps} = \frac{i \alpha}{\pi f_\pi}$, which is clearly at odds with Eq. \ref{MPsEq}. The resolution
of this problem requires a formalism capable of incorporating the effects of dynamical chiral
symmetry breaking, such as described in Refs. \cite{ls,rm}.}.

The total width is evaluated by summing over intermediate states, integrating, and symmetrizing 
appropriately. Form factors and decay constants are computed as described in the preceding sections.
As argued above, form factors 
are evaluated at the point $|\vec q|  = m_{Ps}/2$.  
Table \ref{etaTab} shows the rapid convergence
of the amplitude in the vector principle quantum number $n$  for the quantity
$\frac{4\sqrt{2}}{Q\sqrt{m_{\eta_c}}}{\cal A}_{++}$. Surprisingly, convergence is not so fast for
the $\Upsilon$ system and care must be taken in this case.

\begin{table}[!h]
\caption{Amplitude for $\eta_c\rightarrow\gamma\gamma$ ($10^{-3}$ GeV$^{-1}$).}
\begin{tabular}{c|cc}
\hline
n  & BGS & BGS log \\
\hline
\hline
1 & -211 &  -141 \\
2 &  -34 &  -30  \\
3 &  -10 &  -10  \\
\hline
\hline
\end{tabular}
\label{etaTab}
\end{table}

Table \ref{ccggTab} presents the computed widths for the $\eta_c$, $\eta_c'$, and $\chi_{c0}$ mesons in
a variety of models. The second and third columns compare the predictions of the BGS model with 
and without a running coupling. Use of the running coupling reduces the predictions by approximately
a factor of two, bringing the model into good agreement with experiment.
This is is due, in large part, to the more accurate vector decay constants provided by the BGS+log
model. In comparison, the results of Godfrey and Isgur (labelled GI), which rely on naive factorisation
 supplemented with the {\it ad hoc} pseudoscalar mass dependence discussed above, does not fare so
well for the excited $\eta_c$ transition rate. Similarly a computation using heavy quark effective field
theory (labelled HQ) finds a large $\eta_c'$ rate. Columns 6 and 7 present results computed in the
factorisation approach with nonrelativistic and relativistic wavefunctions respectively.  Columns 8 and 9 
(Munz and  Chao) also use factorisation but compute with the Bethe-Salpeter formalism. The model of 
column 10 (CWV) 
employs factorisation with wavefunctions determined by a two-body Dirac equation. With the exception of
the last model, it appears that
model variation in factorisation approaches can accommodate some, but never all, of the experimental
data, in contrast to the bound state perturbation theory result. However, more and better data are required
before this conclusion can be firm.

\begin{table}[h!]
\caption{Charmonium Two-photon Decay Rates (keV).}
\begin{tabular}{c|cc|ccccccc|c}
\hline
process & BGS & BGS log ($\Lambda=0.25$ GeV) & G\&I\cite{GI} & HQ\cite{LP} & A\&B\cite{AB} & EFG\cite{EFG} & Munz\cite{Munz}& Chao\cite{Chao}& CWV\cite{CWV}& PDG\footnote{The $\eta_c'$ rate is obtained from Ref. \cite{CLEO} and assumes that $Br(\eta_c \to K_SK\pi) = Br(\eta_c' \to K_SK\pi)$. This assumption is supported by the measured
rates for $B \to K\eta_c$ and $K\eta_c'$ as explained in Sect. III.B of Ref. \cite{HHreview}.}
 \\
\hline
\hline
$\eta_c  \rightarrow \gamma\gamma$     &14.2   &7.18     &6.76 & 7.46 & 4.8 & 5.5 & 3.5(4)    & 6-7  & 6.18 & $7.44\pm 2.8$\\
$\eta'_c \rightarrow \gamma\gamma$     &2.59   &1.71     &4.84 & 4.1 & 3.7  & 1.8 & 1.4(3)    &  2   & 1.95 & $1.3 \pm 0.6$ \\
$\eta''_c \rightarrow \gamma\gamma$     &1.78   &1.21     & -- & --  & --   & --  & 0.94(23)  & --   & -- &  -- \\
$\chi_{c0} \rightarrow \gamma\gamma$   &5.77   &3.28     & --  & --  & --   & 2.9 & 1.39(16)  &  --  & 3.34 & $2.63\pm 0.5$ \\
\hline
\hline
\end{tabular}
\label{ccggTab}
\end{table}

\section{Summary and Conclusions} 

We have presented computations of nine charmonium decay constants, eight single quark form
factors, ten radiative transition form factors, and four two-photon decay rates (with an
additional twelve bottomonium decay constants and four two-photon decay rates). Overall, the
agreement with experiment and lattice is impressive.  This level of agreement has been 
achieved with a combination of model building (namely the use of a running coupling
in the traditional Cornell potential model), the incorporation of simple relativistic effects,
and, in the case of two-photon transitions, appropriate computational technique. 

In our view, the combination of the improved methods described above leads to a more
satisfactory quark model phenomenology of dynamical properties of mesons. Specifically, form factor
momentum 
rescaling constants, artificial energy dependence in decay constant integrals,  and {\it ad hoc} phase
space redefinitions in two-photon decays are no longer required. Furthermore, we have demonstrated
that ambiguities due to the noncovariance of the nonrelativistic constituent quark model can be 
expected to give
rise to theoretical uncertainties on the order of 10\% and thus need not invalidate the method for
processes with sufficiently low recoil momenta.

Nevertheless, there are strong hints that flaws remain in the constituent quark model.
First, it appears to be difficult to maintain the excellent agreement of the  nonrelativistic
phenomenological spectrum with experiment when a running coupling is employed. Second, 
predicted
decay constants of highly excited vectors appear to be too large with
respect to experiment. Thus the short distance strength in the wavefunctions is not dropping
sufficiently rapidly with principle quantum number. Similarly, the large $\psi(3770)$ decay 
constant is difficult to reconcile with the model presented here.
These difficulties imply that there is additional room for improved hadronic model building. 
Obvious possibilities include relativistic models and the incorporation of Fock sector mixing.

Overall, the success of the computations presented here fosters confidence in the model and 
techniques and we look forward to applying them to other processes of interest (such as
electroweak processes relevant to heavy meson decays).

\acknowledgments

We are grateful to Ted Barnes, Stan Brodsky, Frank Close, and Jo Dudek for discussions.
We also thank the authors of Ref. \cite{JLlatt} for making their results available to us.
This work is supported by the U.S. Department of Energy under contract DE-FG02-00ER41135.

\appendix

\section{Model Parameters}
\label{modelsApp}

Charmonia are described with the Hamiltonian of Eqs. \ref{vcEq} and \ref{vsdEq} and the parameters
determined in Ref. \cite{BGS} by fitting the known charmonium specrtum: $m_c = 1.4794$ GeV, $\alpha_c = \alpha_H = 0.5461$, $\sigma = 1.0946$ GeV,
and $b = 0.1425$ GeV$^2$. No constant is included.  The `BGS+log' model retains the same parameters 
as above with the exception that the Coulomb strong coupling constant is replaced with the running
coupling of Eq. \ref{alRunEq}. In this case we set $\alpha_0 = \alpha_H = 0.5461$.

The bottomonium parameters were obtained by fitting the model of Eqs. \ref{vcEq} and \ref{vsdEq} to the known bottomonium spectrum. The results were $m_b = 4.75$ GeV,
$\alpha_C = \alpha_H =0.35$, $b=0.19$ GeV$^2$, and $\sigma=0.897$ GeV.

\section{Bottomonium Properties}
\label{bottomApp}

Predicted bottomonium spectra, decay constants, and two-photon decay rates are presented here.
All computations we performed as for charmonia.

\begin{table}[!h]
\caption{Bottomonium Spectrum (GeV).}
\begin{center}
\begin{tabular}{c|cccc}
\hline
Meson & C+L & C+L log  & C+L log & PDG \\
      &     & $\Lambda = 0.4$ GeV & $\Lambda = 0.25$ GeV & \\
\hline
\hline
$\eta_b$                       &9.448   &9.490   &9.516    &\\
$\eta_b'$                      &10.006  &10.023  &10.033   &\\
$\eta_b''$                     &10.352  &10.365  &10.372   &\\
$\Upsilon$                     &9.459   &9.500   &9.525    &$9.4603\pm 0.00026$\\
$\Upsilon'$                    &10.009  &10.026  &10.036   &$10.02326\pm 0.00031$\\
$\Upsilon''$                   &10.354  &10.367  &10.374   &$10.3552\pm 0.0005$\\
$\chi_{b0}$                       &9.871   &9.873   &9.879    &$9.8599\pm 0.001$\\
$\chi_{b0}'$                      &10.232  &10.235  &10.239   &$10.2321\pm 0.0006$\\
$\chi_{b0}''$                     &10.522  &10.525  &10.529   &\\
$\chi_{b1}$                       &9.897   &9.900   &9.904    &$9.8927\pm 0.0006$\\
$\chi_{b1}'$                      &10.255  &10.257  &10.260   &$10.2552\pm 0.0005$\\
$\chi_{b1}''$                     &10.544  &10.546  &10.548   &\\
$\chi_{b2}$                       &9.916   &9.917   &9.921    &$9.9126\pm 0.0005$\\
$\chi_{b2}'$                      &10.271  &10.272  &10.275   &$10.2685\pm 0.0004$\\
$\chi_{b2}''$                     &10.559  &10.560  &10.563   &\\
\hline
\hline
\end{tabular}
\end{center}
\end{table}

\begin{table}[!h]
\caption{Bottomonium Decay Constants (MeV).}
\begin{center}
\begin{tabular}{c|ccccc}
\hline
Meson & C+L NonRel & C+L Rel & C+L log  & C+L log & experiment \\
      &            &         & $\Lambda = 0.4$ GeV  & $\Lambda = 0.25$ GeV & \\
\hline
\hline
$\eta_b$                       &979    &740   &638   &599     &\\
$\eta_b'$                      &644    &466   &423   &411     &\\
$\eta_b''$                     &559    &394   &362   &354     &\\
$\Upsilon$                     &963    &885   &716   &665     &$708\pm 8$\\
$\Upsilon'$                    &640    &581   &495   &475     &$482\pm 10$\\
$\Upsilon''$                   &555    &501   &432   &418     &$346\pm 50$\\
$\Upsilon'''$                  &512    &460   &400   &388     &$325\pm 60$\\
$\Upsilon^{(4)}$                &483    &431   &377   &367      &$369\pm 93$\\
$\Upsilon^{(5)}$                 &463    &412   &362   &351     &$240\pm 61$\\
$\chi_{b1}$                       &186    &150    &142    &136   &\\
$\chi_{b1}'$                      &205    &160   &152   &147  &\\
$\chi_{b1}''$                     &215    &164   &157   &152  &\\
\hline
\hline
\end{tabular}
\end{center}
\end{table}

\begin{table}[h!]
\caption{Bottomonium Two-photon Decay Rates (keV)}
\begin{center}
\begin{tabular}{c|cccc}
\hline
process & C+L & C+L log & GI & experiment \\
        &     & $\Lambda = 0.25$ GeV & & \\ 
\hline
\hline
$\eta_b  \rightarrow \gamma\gamma$     &0.45    &0.23    & 0.38  & -- \\
$\eta_b' \rightarrow \gamma\gamma$     &0.11    &0.07    & --    & --  \\
$\eta_b'' \rightarrow \gamma\gamma$    &0.063   &0.040   & --    & --  \\
$\chi_{b0} \rightarrow \gamma\gamma$   &0.126   &0.075   & --    & --  \\
\hline
\hline
\end{tabular}
\end{center}
\end{table}

\section{Decay Constants}
\label{DecayConstantsApp}

Decay constant definitions and quark model expressions for vector, scalar, pseudoscalar, axial, and 
$h_c$
meson decay constants are presented here.

\subsection{Vector Decay Constant}

The decay constant $f_V$ of the vector meson is defined as

\begin{equation}
m_Vf_V\epsilon^{\mu}=\langle0|\bar{\Psi}\gamma^{\mu}\Psi|V\rangle
\end{equation}
where $m_V$ is the vector meson mass, $\epsilon^{\mu}$ is its polarization vector, $|V\rangle$ is the vector meson state.  The decay constant has been extracted from leptonic decay rates with the aid of the
following:
\begin{equation}
\Gamma_{V\rightarrow e^{+}e^{-}}=\frac{e^4Q^2f^2_V}{12\pi m_V}=\frac{4\pi\alpha^2}{3}\frac{Q^2f^2_V}{m_V}.
\end{equation}

Following the method described in the text yields the quark model 
vector meson decay constant:
\begin{equation}\label{relfV}
f_V=\sqrt{\frac{3}{m_V}}\int\frac{d^3k}{(2\pi)^3} \Phi(\vec{k})
\sqrt{1+\frac{m_q}{E_k}} \sqrt{1+\frac{m_{\bar{q}}}{E_{\bar{k}}}}
\left(1+\frac{k^2}{3(E_k+m_q)(E_{\bar{k}}+m_{\bar{q}})}\right)
\end{equation}

The nonrelativistic limit of this yields the well-known proportinality of the decay constant to the
wavefunction at the origin:
\begin{equation}
f_V=2\sqrt{\frac{3}{m_V}}\int\frac{d^3k}{(2\pi)^3} \Phi(\vec{k})
=2\sqrt{\frac{3}{m_V}} \tilde{\Phi}(r=0).
\end{equation}

\subsection{Pseudoscalar Decay Constant}

The decay constant $f_P$ of a pseudoscalar meson is defined by

\begin{equation}
p^{\mu}f_P= i \langle0|\bar{\Psi}\gamma^{\mu}\gamma^5\Psi|P\rangle
\end{equation}
where $p^{\mu}$ is the meson momentum and $|P\rangle$ is the pseudoscalar meson state.
The pseudoscalar decay rate is then
\begin{equation}
\Gamma_{P\rightarrow l^{+}{\nu}_l}=\frac{G_F^2}{8\pi} |V_{q\bar{q}}|^2 f_P^2 m_l^2 m_P \left(1-\frac{m_l^2}{m_P^2}\right)^2.
\end{equation}

The quark model expression for the decay constant is
\begin{equation}
f_P=\sqrt{\frac{3}{m_P}}\int\frac{d^3k}{(2\pi)^3}
\sqrt{1+\frac{m_q}{E_k}}\sqrt{1+\frac{m_{\bar{q}}}{E_{\bar{k}}}}
\left(1- \frac{k^2}{(E_k+m_q)(E_{\bar{k}}+m_{\bar{q}})} \right)
\Phi(\vec{k}).
\end{equation}

In the nonrelativistic limit this reduces to the same expression as the vector decay constant.

\subsection{Scalar Decay Constant}

The decay constant $f_S$ of the scalar meson is defined by

\begin{equation}
p^{\mu}f_S=\langle 0|\bar{\Psi}\gamma^{\mu}\Psi|S\rangle,
\end{equation}
which yields the quark model result:
\begin{equation}
f_S=\sqrt{\frac{3}{m_S}}\frac{\sqrt{4\pi}}{(2\pi)^3}
\int k^3 dk\, \sqrt{1+\frac{m_q}{E_k}}\sqrt{1+\frac{m_{\bar{q}}}{E_{\bar{k}}}}
\left(\frac{1}{E_{\bar{k}}+m_{\bar{q}}}-\frac{1}{E_k+m_q} \right) R(k).
\end{equation}
Here and in the following,  $R$ is the radial wavefunction defined by $\Phi(k) = Y_{lm}R(k)$ with
$\int \frac{d^3k}{(2\pi)^3} |\Phi|^2 = 1$.


\subsection{Axial Vector Decay Constant}

The decay constant $f_A$ of the axial vector meson is defined as

\begin{equation}
\epsilon^{\mu}f_Am_A=\langle 0|\bar{\Psi}\gamma^{\mu}\gamma^5\Psi|A\rangle
\end{equation}
where $\epsilon^{\mu}$ is the meson polarization vector, $m_A$ is its mass and $|A\rangle$ is the axial vector meson state. The quark model decay constant is thus
\begin{equation}
f_A=-\sqrt{\frac{2}{m_A}}\frac{\sqrt{4\pi}}{(2\pi)^3}
\int k^3 dk\,\sqrt{1+\frac{m_q}{E_k}}\sqrt{1+\frac{m_{\bar{q}}}{E_{\bar{k}}}}
\left(\frac{1}{E_{\bar{k}}+m_{\bar{q}}}+\frac{1}{E_k+m_q} \right) R(k).
\end{equation}

\subsection{$h_c$ Decay Constant}

The decay constant $f_{A'}$ of the $^1P_1$ state meson is defined by:

\begin{equation}
\epsilon^{\mu}f_{A'}m_{A'}=\langle 0|\bar{\Psi}\gamma^{\mu}\gamma^5\Psi|^1P_1\rangle
\end{equation}
where $\epsilon^{\mu}$ is the meson polarization vector, $m_{A'}$ is its mass and $|^1P_1\rangle$ is its state.
 The resulting quark model decay constant is given by
\begin{equation}
f_{A'}=\frac{1}{\sqrt{m_{A'}}}\frac{\sqrt{4\pi}}{(2\pi)^3}
\int k^3 dk \,\sqrt{1+\frac{m_q}{E_k}}\sqrt{1+\frac{m_{\bar{q}}}{E_{\bar{k}}}}
\left(\frac{1}{E_{\bar{k}}+m_{\bar{q}}}-\frac{1}{E_k+m_q} \right) R(k).
\end{equation}

\section{Form Factors}
\label{ffApp}

A variety of Lorentz invariant multipole decompositions (see Ref. \cite{JLlatt}) and quark 
model expressions for
these multipoles are presented in the following. 

Each transition form-factor is normally a sum of two terms corresponding to the coupling of the external current to the quark and antiquark. For $c\bar{c}$-mesons these two terms are equal to each other, so in the following we only present formulas corresponding to the single quark coupling. In general both terms have to be calculated.

\subsection{Pseudoscalar Form Factor}

The most general Lorentz covariant decomposition for the electromagnetic transition matrix element between two pseudoscalars is:
\begin{equation}
\langle P_2(p_2)|\bar{\Psi}\gamma^{\mu}\Psi|P_1(p_1)\rangle=f(Q^2)(p_2+p_1)^{\mu}+g(Q^2)(p_2-p_1)^{\mu}
\end{equation}

To satisfy time-reversal invariance the form-factors $f(Q^2)$ and $g(Q^2)$ have to be real.
The requirement that the vector current is locally conserved gives a relation between two form-factors:
\begin{equation}
g(Q^2)=f(Q^2)\frac{M^2_2-M^2_1}{Q^2}.
\end{equation}
Thus the matrix element can be written as:
\begin{equation}
\langle P_2(p_2)|\bar{\Psi}\gamma^{\mu}\Psi|P_1(p_1)\rangle=f(Q^2)\left((p_2+p_1)^{\mu}-\frac{M^2_2-M^2_1}{q^2}(p_2-p_1)^{\mu}\right)
\end{equation}
In case of two identical pseudoscalars the second term vanishes.

Computing with the temporal component of the current in the quark model formalism yields (for $c\bar{c}$-mesons)
\begin{eqnarray}
f(Q^2)&=&\frac{\sqrt{M_1E_2}}{(E_2+M_1)-\frac{M^2_2-M^2_1}{q^2}(E_2-M_1)}\nonumber\\
&\times&\int \frac{d^3k}{(2\pi)^3}
\Phi(\vec{k})
\Phi^*\left(\vec{k}+\frac{\vec{q}}{2}\right)
\sqrt{1+\frac{m_q}{E_k}}\sqrt{1+\frac{m_q}{E_{k+q}}}
\left(1+\frac{(\vec{k}+\vec{q})\cdot\vec{k}}
{(E_k+m_q)(E_{k+q}+m_q)}\right)
\end{eqnarray}

In case of identical pseudoscalars in the non-relativistic approximation the formula above simplifies to
\begin{equation}\label{ff0}
f(Q^2)=\frac{2\sqrt{M_1E_2}}{E_2+M_1}
\int \frac{d^3k}{(2\pi)^3}
\Phi(\vec{k})
\Phi^*\left(\vec{k}+\frac{\vec{q}}{2}\right).
\end{equation}

Similar expressions occur when the computation is made with the spatial components of
the electromagnetic current:
\begin{eqnarray}
f(Q^2)\!\!\!&=&\!\!\!\frac{\sqrt{M_1E_2}}{1-\frac{M^2_2-M^2_1}{q^2}}
\,\frac{\vec{q}}{|\vec{q}|^2}\cdot\!
\int \frac{d^3k}{(2\pi)^3}
\Phi(\vec{k})
\Phi^*\left(\vec{k}+\frac{\vec{q}}{2}\right)
\sqrt{1+\frac{m_q}{E_k}}\sqrt{1+\frac{m_q}{E_{k+q}}}
\left(\frac{\vec{k}}{E_k+m_q}+\frac{\vec{k}+\vec{q}}
{E_{k+q}+m_q}\right)
\label{chiSpatialEq}
\end{eqnarray}

In this case the nonrelativistic approximation for the single quark form factor is
\begin{eqnarray}
f(Q^2)=\frac{\sqrt{M_1E_2}}{m|\vec{q}|^2}
\vec{q}\cdot\int \frac{d^3k}{(2\pi)^3}
\Phi(\vec{k})
\Phi^*\left(\vec{k}+\frac{\vec{q}}{2}\right)(2\vec{k}+\vec{q}).
\end{eqnarray}

%
%
Covariance requires the same expression for the temporal and spatial form factors. Comparing the 
formula above to the expression for the temporal form factor (\ref{ff0}) shows that covariance is 
recovered in the nonrelativistic and weak coupling limits (where $M_1+M_2 \to 4m$).

\subsection{Vector Form Factors}

The most general Lorentz covariant decomposition for the electromagnetic transition matrix element between two identical vectors is:
\begin{eqnarray}
\langle V(p_2)|\bar{\Psi}\gamma^{\mu}\Psi|V(p_1)\rangle=
-(p_1+p_2)^{\mu}\left[G_1(Q^2) (\epsilon^*_2 \cdot \epsilon_1)
+ \frac{G_3(Q^2)}{2m^2_V} (\epsilon^*_2 \cdot p_1) (\epsilon_1 \cdot p_2) \right]\nonumber\\
+G_2(Q^2)\left[\epsilon^{\mu}_1 (\epsilon^*_2 \cdot p_1) + \epsilon^{\mu*}_2 (\epsilon_1 \cdot p_2) \right]
\end{eqnarray}
These form-factors are related to the standard charge, magnetic dipole and quadrupole multipoles by
\begin{eqnarray}
G_C&=&\left(1+\frac{2}{3}\eta\right)G_1-\frac{2}{3}\eta G_2+\frac{2}{3}\eta(1+\eta)G_3\nonumber\\
G_M&=&G_2\nonumber\\
G_Q&=&G_1-G_2+(1+\eta)G_3
\label{G3Eq}
\end{eqnarray}
where $\eta=\frac{Q^2}{4m^2_q}$.

Quark model expressions for these are:
\begin{eqnarray}
G_2(Q^2)=-\frac{\sqrt{m_VE_2}}{|\vec{q}|^2}
\int \frac{d^3k}{(2\pi)^3}
\Phi(\vec{k})
\Phi^*\left(\vec{k}+\frac{\vec{q}}{2}\right)
\sqrt{1+\frac{m_q}{E_k}}\sqrt{1+\frac{m_q}{E_{k+q}}}
\left(
 \frac{\vec{k}\cdot\vec{q}}{E_k+m_q}-\frac{\vec{k}\cdot\vec{q}+|\vec{q}|^2}{E_{k+q}+m_q}\right)
\label{GMEq}
\end{eqnarray}
and
\be
G_1(Q^2)=\frac{\sqrt{m_VE_2}}{m_V+E_2}
\int \frac{d^3k}{(2\pi)^3}
\Phi(\vec{k})
\Phi^*\left(\vec{k}+\frac{\vec{q}}{2}\right)
\sqrt{1+\frac{m_q}{E_k}}\sqrt{1+\frac{m_q}{E_{k+q}}}
\left(1+\frac{(\vec{k}+\vec{q})\cdot\vec{k}}
{(E_k+m_q)(E_{k+q}+m_q)}\right)
\ee
or
\be
G_1(Q^2)=\frac{\sqrt{m_VE_2}}{|\vec{q}|^2}
\int \frac{d^3k}{(2\pi)^3}
\Phi(\vec{k})
\Phi^*\left(\vec{k}+\frac{\vec{q}}{2}\right)
\sqrt{1+\frac{m_q}{E_k}}\sqrt{1+\frac{m_q}{E_{k+q}}}
\left(\frac{\vec{k}\cdot\vec{q}}{E_k+m_q}+\frac{\vec{k}\cdot\vec{q}+|\vec{q}|^2}{E_{k+q}+m_q}\right).
\ee

$G_3$ can be expressed in terms of $G_1$ and $G_2$ in two different ways:
\begin{equation}
G_3=\frac{2m^2_V}{|\vec{q}|^2}\left(1-\frac{E_2}{m_V}\right)G_1+\frac{2m_V}{E_2+m_V}G_2
\end{equation}
or
\begin{equation}
G_3=\frac{2m_V(m_V-E_2)}{|\vec{q}|^2}(G_1-G_2).
\end{equation}
One can establish that $G_3\rightarrow G_2-G_1$ as $|\vec{q}|\rightarrow 0$ from either equation.

\subsection{Scalar Form Factor}

The most general Lorentz covariant decomposition for the electromagnetic transition matrix element between two scalars is:
\begin{equation}
\langle S_2(p_2)|\bar{\Psi}\gamma^{\mu}\Psi|S_1(p_1)\rangle=f(Q^2)(p_2+p_1)^{\mu}+g(Q^2)(p_2-p_1)^{\mu}.
\end{equation}

As with pseudoscalars, this can be written as
\begin{equation}
\langle S_2(p_2)|\bar{\Psi}\gamma^{\mu}\Psi|S_1(p_1)\rangle=f(Q^2)\left((p_2+p_1)^{\mu}-\frac{M^2_2-M^2_1}{q^2}(p_2-p_1)^{\mu}\right).
\end{equation}

In the case of identical scalars the quark model calculation gives
\begin{eqnarray}
f(Q^2)=\frac{\sqrt{M_1E_2}}{E_2+M_1}
\int \frac{d^3k}{(2\pi)^3}
\Phi(\vec{k})
\Phi^*\left(\vec{k}+\frac{\vec{q}}{2}\right)
\sqrt{1+\frac{m_q}{E_k}}\sqrt{1+\frac{m_q}{E_{k+q}}}
\left(1+\frac{(\vec{k}+\vec{q})\cdot\vec{k}}{(E_k+m_q)(E_{k+q}+m_q)}\right).
\end{eqnarray}

In the nonrelativistic limit this reduces to

\begin{equation}
f(Q^2)=
\int \frac{d^3k}{(2\pi)^3}
\Phi(\vec{k})
\Phi^*\left(\vec{k}+\frac{\vec{q}}{2}\right).
\end{equation}

%

\subsection{Vector-Pseudoscalar Transition Form Factor}

The most general Lorentz covariant decomposition for the electromagnetic transition matrix element between vector and pseudoscalar is:
\begin{equation}
\langle P(p_P)|\bar{\Psi}\gamma^{\mu}\Psi|V(p_V)\rangle=iF(Q^2)\epsilon^{\mu\nu\alpha\beta}(\epsilon_{M_V})_{\nu}(p_V)_{\alpha}(p_P)_{\beta}.
\end{equation}
Computing with the spatial components of the current then gives

\be
F(Q^2) = -\sqrt{\frac{E_P}{m_V}}
\,\frac{1}{|\vec{q}|^2}
\int \frac{d^3k}{(2\pi)^3}
\Phi_V(\vec{k})
\Phi_P^*\left(\vec{k}+\frac{\vec{q}}{2}\right)
\sqrt{1+\frac{m_q}{E_k}}\sqrt{1+\frac{m_q}{E_{k+q}}}
\left(\frac{\vec{k}\cdot\vec{q}}{E_k+m_q}-\frac{\vec{k}\cdot\vec{q}+|\vec{q}|^2}
{E_{k+q}+m_q}\right).
\ee

In the nonrelativistic approximation in zero recoil limit $\vec{q}\rightarrow 0$ this reduces to
\begin{equation}
F(Q^2)|_{\vec{q}\rightarrow 0}=\frac{1}{m_q}\sqrt{\frac{m_P}{m_V}}.
\end{equation}

\subsection{Scalar-Vector Transition Form Factors}

The most general Lorentz covariant decomposition for the electromagnetic transition matrix element between scalar ($^3P_0$) meson state and vector ($^3S_1$) is
\begin{eqnarray}
\langle V(p_V)|\bar{\Psi}\gamma^{\mu}\Psi|S(p_S)\rangle=
\Omega^{-1}(Q^2)\Bigg( E_1(Q^2)\left[ \Omega(Q^2)\epsilon^{*\mu}_{M_V}-\epsilon^*_{M_V}\cdot p_S
(p^{\mu}_V p_V\cdot p_S-m^2_V p^{\mu}_S)\right]\Bigg.\nonumber\\
\left.+\frac{C_1(Q^2)}{\sqrt{Q^2}}m_V \epsilon^*_{M_V}\cdot p_S\left[p_V\cdot p_S(p_V+p_S)^{\mu}-m^2_S p^{\mu}_V-m^2_V p^{\mu}_S\right]\right)
\end{eqnarray}
where $\Omega(Q^2) \equiv (p_V\cdot p_S)^2-m^2_Vm^2_S=\frac{1}{4}\left[(m_V-m_S)^2-Q^2\right]\left[(m_V+m_S)^2-Q^2\right]$, and takes the simple value $m^2_s|\vec{q}|^2$ in the rest frame of a decaying scalar.

$E_1$ contributes to the amplitude only in the case of transverse photons, while $C_1$ contributes 
only for longitudinal photons. Quark model expressions for the multipole form factors are

\begin{equation}
C_1(Q^2)
= -2\frac{\sqrt{Q^2}}{|\vec{q}|}\frac{\sqrt{E_Vm_S}}{4\pi}
\int \frac{d^3k}{(2\pi)^3}
R_S(\vec{k})R_V\left(\vec{k}+\frac{\vec{q}}{2}\right)
\sqrt{1+\frac{m_q}{E_k}}\sqrt{1+\frac{m_q}{E_{k+q}}}
\left(\cos{\Theta}+\frac{k^2+|\vec{k}|\cdot|\vec{q}|}
{(E_k+m_q)(E_{k+q}+m_q)}\right)\
\end{equation}

\begin{eqnarray}
C_1(Q^2)
&=& 2\frac{\sqrt{E_Vm_S}}{4\pi}\frac{\sqrt{Q^2}}{|\vec{q}|}
\int \frac{d^3k}{(2\pi)^3}
R_S(\vec{k})
R_V\left(\vec{k}+\frac{\vec{q}}{2}\right)
\sqrt{1+\frac{m_q}{E_k}}\sqrt{1+\frac{m_q}{E_{k+q}}}\nonumber\\
&&\times\left(\frac{k}{E_k+m_q}+\frac{q\cos{\Theta}}{E_{k+q}+m_q}
+\frac{k\cos{2\Theta}}{E_{k+q}+m_q}\right).
\end{eqnarray}
The first(second) expression for $C_1(Q^2)$ is calculated from the temporal(spatial) matrix element of the current.

\begin{eqnarray}
E_1(Q^2)=-2\frac{\sqrt{E_Vm_S}}{4\pi}
\int \frac{d^3k}{(2\pi)^3}
R_S(\vec{k})
R_V\left(\vec{k}+\frac{\vec{q}}{2}\right)
\sqrt{1+\frac{m_q}{E_k}}\sqrt{1+\frac{m_q}{E_{k+q}}}\left[\frac{k}{E_k+m_q}
-\frac{k\cos{\Theta}+q}{E_{k+q}+m_q}\right]\nonumber.
\end{eqnarray}

\subsection{$h_{c1}$-Pseudoscalar Transition Form Factor}

The most general Lorentz covariant decomposition for the electromagnetic transition matrix element 
between $^1P_1$ meson state and pseudoscalar ($^1S_0$) is

\begin{eqnarray}
\langle P(p_P)|\bar{\Psi}\gamma^{\mu}\Psi|A(p_A)\rangle=\Omega^{-1}(Q^2)\Bigg( E_1(Q^2)\left[ 
\Omega(Q^2)\epsilon^{\mu}_{M_L}-\epsilon_{M_L}\cdot p_P(p^{\mu}_A p_A\cdot p_P-m^2_A p^{\mu}_P)
\right]\Bigg.\nonumber\\
\left.+\frac{C_1(Q^2)}{\sqrt{Q^2}}m_A \epsilon_{M_L}\cdot p_P\left[p_A\cdot p_P(p_A+p_P)^{\mu}-m^2_P 
p^{\mu}_A-m^2_A p^{\mu}_P\right]\right).
\end{eqnarray}

Quark model expressions for the form factors are 
\be
E_1(Q^2)=\frac{\sqrt{3m_AE_P}}{8\pi}\,
\int \frac{d^3k}{(2\pi)^3}
R_A(\vec{k})
R_P\left(\vec{k}+\frac{\vec{q}}{2}\right)
\sqrt{1+\frac{m_q}{E_k}}\sqrt{1+\frac{m_q}{E_{k+q}}}
k\sin^2{\Theta}
\left(\frac{1}{E_k+m_q}+\frac{1}{E_{k+q}+m_q}\right)
\ee
and

\be
C_1(Q^2) = -\frac{\sqrt{3m_AE_P}}{4\pi}\frac{\sqrt{Q^2}}{|\vec{q}|}\,
\int \frac{d^3k}{(2\pi)^3}
R_A(\vec{k})
R_P\left(\vec{k}+\frac{\vec{q}}{2}\right)
\sqrt{1+\frac{m_1}{E_k}}\sqrt{1+\frac{m_2}{E_{k+q}}}
\cos{\Theta}
\left(1+\frac{k^2+kq\cos{\Theta}}{(E_k+m_q)(E_{k+q}+m_q)}\right).
\ee

\subsection{Axial Vector - Vector Transition Form Factor}

The most general Lorentz covariant decomposition for the electromagnetic transition matrix element 
between axial vector ($^3P_1$) meson state and vector ($^3S_1$) is

\begin{eqnarray}
&&\langle V(p_V)|\bar{\Psi}\gamma^{\mu}\Psi|A(p_A)\rangle=\frac{i}{4\sqrt{2}\Omega(Q^2)}\epsilon^{\mu\nu\rho\sigma}(p_A-p_V)_{\sigma}\times\nonumber\\
&\times& \Bigg[ E_1(Q^2)(p_A+p_V)_{\rho}\bigg(2m_A[\epsilon_{M_A}\cdot p_V](\epsilon^*_{M_V})_{\nu}
+2m_V[\epsilon^*_{M_V}\cdot p_A](\epsilon_{M_A})_{\nu}\bigg)\nonumber\\
&&+M_2(Q^2)(p_A+p_V)_{\rho}\bigg(2m_A[\epsilon_{M_A}\cdot p_V](\epsilon^*_{M_V})_{\nu}
-2m_V[\epsilon^*_{M_V}\cdot p_A](\epsilon_{M_A})_{\nu}\bigg)\nonumber\\
&&+\frac{C_1(Q^2)}{\sqrt{Q^2}}\bigg(-4\Omega(Q^2)(\epsilon_{M_A})_{\nu} (\epsilon^*_{M_V})_{\rho}\nonumber\\
&&+(p_A+p_V)_{\rho}\bigg[
(m^2_A-m^2_V+Q^2)[\epsilon_{M_A}\cdot p_V](\epsilon^*_{M_V})_{\nu}+
(m^2_A-m^2_V-Q^2)[\epsilon^*_{M_V}\cdot p_A](\epsilon_{M_A})_{\nu}
\bigg]\bigg)\Bigg]
\end{eqnarray}

Quark model expressions for the form factors are 
\be
E_1(Q^2)=-\frac{\sqrt{3m_AE_V}}{8\pi}\,
\int \frac{d^3k}{(2\pi)^3}
R_A(\vec{k})
R_V\left(\vec{k}+\frac{\vec{q}}{2}\right)
\sqrt{1+\frac{m_q}{E_k}}\sqrt{1+\frac{m_q}{E_{k+q}}}
\left(\frac{k(3-\cos^2{\Theta})}{E_k+m_q}+\frac{k(1-3\cos^2{\Theta})-2q\cos{\Theta}}{E_{k+q}+m_q}\right)
\ee

\be
M_2(Q^2)=-\frac{\sqrt{3m_AE_V}}{8\pi}\,
\int \frac{d^3k}{(2\pi)^3}
R_A(\vec{k})
R_V\left(\vec{k}+\frac{\vec{q}}{2}\right)
\sqrt{1+\frac{m_q}{E_k}}\sqrt{1+\frac{m_q}{E_{k+q}}}
\left(\frac{k(1-3\cos^2{\Theta})}{E_k+m_q}-\frac{k(1-3\cos^2{\Theta})+2q\cos{\Theta}}{E_{k+q}+m_q}\right)
\ee

and

\be
C_1(Q^2) = \frac{\sqrt{3m_AE_V}}{2\pi}\frac{\sqrt{Q^2}}{|\vec{q}|}\,
\int \frac{d^3k}{(2\pi)^3}
R_A(\vec{k})
R_V\left(\vec{k}+\frac{\vec{q}}{2}\right)
\sqrt{1+\frac{m_q}{E_k}}\sqrt{1+\frac{m_q}{E_{k+q}}}
\left(\cos{\Theta}+\frac{k^2\cos{\Theta}+\frac{1}{2}kq(1+\cos^2{\Theta})}
{(E_k+m_q)(E_{k+q}+m_q)}\right).
\ee

\end{document}